\documentclass[%
 aip,
 amsmath,amssymb,
preprint,%
]{revtex4-1}
\usepackage{graphicx} 
\usepackage{float}
\usepackage{tabularx}
\usepackage[hidelinks]{hyperref}
\usepackage{physics}
\usepackage{braket}
\usepackage{tikz}
\usepackage{xurl}

\usepackage{chngcntr}

\newcommand{\Ms}{M_\textrm{S}}
\newcommand{\Mo}{M_\textrm{O}}
\newcommand{\dt}{\delta t}
\newcommand{\betacd}{\beta_{\textrm{CD}}}
\newcommand{\thetacd}{\theta_{\textrm{CD}}}
\DeclareMathOperator\arctanh{arctanh}
\newcommand{\cminv}{\textrm{cm}^{-1}}

\usepackage{tikz}
\usepackage{ifthen}
\usepackage{xstring}

\newenvironment{diagram}
{
\begin{tikzpicture}[baseline = (X.base),every node/.style={scale=1.},scale=.7]
}
{
\end{tikzpicture}
}
\newcommand{\MpsCircle}[4]{
\IfSubStr{#4}{u}{ \draw (#1, #2+0.5) -- (#1, #2+1);  }{  }
\IfSubStr{#4}{d}{ \draw (#1, #2-0.5) -- (#1, #2-1); }{  }
\IfSubStr{#4}{l}{ \draw (#1-0.5, #2) -- (#1-1, #2); }{  }
\IfSubStr{#4}{r}{ \draw (#1+0.5, #2) -- (#1+1, #2); }{  }
\IfSubStr{#4}{b}{ \draw (#1, #2) -- (#1+0.5, #2+0.866); \draw (#1, #2) -- (#1-0.5, #2+0.866);}{  }
\draw[fill=white] (#1, #2) circle (.5);
\draw (#1, #2) node {#3};
}

\newcommand{\rev}[1]{{\color{black} #1}}

\begin{document}
\title{Optimal Tree Tensor Network Operators for Tensor Network Simulations: Applications to Open Quantum Systems}
\author{Weitang Li}
\email{liwt31@gmail.com}
\affiliation{School of Science and Engineering, The Chinese University of Hong Kong, Shenzhen, 518172, P. R. China.}
\affiliation{Tencent Quantum Lab, Tencent, Shenzhen, 518057, P. R. China}

\author{Jiajun Ren}
\affiliation{MOE Key Laboratory of Theoretical and Computational
Photochemistry, College of Chemistry, Beijing Normal University,
Beijing, 100875, P. R. China}
\author{Hengrui Yang}
\affiliation{MOE Key Laboratory of Organic OptoElectronics and Molecular Engineering, Department of Chemistry, Tsinghua University, 100084 Beijing, P. R. China}
\author{Haobin Wang}
\affiliation{Department of Chemistry, University of
Colorado Denver, Denver, Colorado 80217-3364, United
States}
\author{Zhigang Shuai}
\affiliation{School of Science and Engineering, The Chinese University of Hong Kong, Shenzhen, 518172, P. R. China.}
\affiliation{MOE Key Laboratory of Organic OptoElectronics and Molecular Engineering, Department of Chemistry, Tsinghua University, 100084 Beijing, P. R. China}


\begin{abstract}
Tree tensor network states (TTNS) decompose the system wavefunction to the product of low-rank tensors based on the tree topology, 
serving as the foundation of the multi-layer multi-configuration time-dependent Hartree (ML-MCTDH) method.
In this work, we present an algorithm that automatically constructs the optimal and exact tree tensor network operators (TTNO) for any sum-of-product symbolic quantum operator.
The construction is based on the minimum vertex cover of a bipartite graph. 
With the optimal TTNO, we simulate open quantum systems such as spin relaxation dynamics in the spin-boson model and charge transport in molecular junctions.
In these simulations, the environment is treated as discrete modes and its wavefunction is evolved on equal footing with the system. 
We employ the Cole-Davidson spectral density to model the glassy phonon environment, and incorporate temperature effects via thermo field dynamics.
Our results show that the computational cost scales linearly with the number of discretized modes, demonstrating the efficiency of our approach.
\end{abstract}

\maketitle

\section{Introduction}
Tensor network algorithms have emerged as a highly effective numerical approach for studying open quantum systems. 
These algorithms decompose the combined system and bath wavefunction into a product of low-rank tensors, providing both efficiency and flexibility.
The multilayer multiconfiguration time-dependent Hartree method~\cite{meyer1990multi, beck2000multiconfiguration, wang2003multilayer} (ML-MCTDH), which utilizes tree tensor network states (TTNS) ansatz~\cite{shi2006classical,nakatani2013efficient, larsson2024tensor},
has been successfully employed to investigate a wide range of problems.
These include non-adiabatic chemical reactions in the condensed phase~\cite{thoss2006quantum, wang2007quantum, westermann2011photodissociation},
charge and heat transfer through molecular junctions~\cite{velizhanin2008heat, wang2011numerically, wang2013numerically},
relaxation dynamics of a spin coupled to various types of baths~\cite{wang2008coherent, wang2012dynamics}
and energy transfer in biological molecules~\cite{schulze2016multi}.
TTNS has also found applications in \textit{ab initio} quantum chemistry~\cite{nakatani2013efficient,gunst2018t3ns, li2021expressibility}, computation of vibrational eigenstates~\cite{larsson2019computing}, solving hierarchical
equations of motion~\cite{yan2021efficient, ke2023tree}, among others~\cite{guo2012critical, schroder2019tensor}.
Another branch of tensor network algorithms is time-dependent density matrix renormalzation group (TD-DMRG), which is based on matrix product states (MPS)~\cite{schollwock2011density, ren2022time}.
MPS is a specific form of tensor networks where the low-rank tensors are arranged in a one-dimensional chain.
Originally designed to address problems related to the ground state or low-lying excited states~\cite{white1992density, white1993density, shuai1998linear},
DMRG has recently been extended to the time-dependent domain~\cite{paeckel2019time, ren2022time}.
This extension has enabled the simulation of
ultra-fast exciton dynamics in organic materials~\cite{Ma16, Mannouch181, Ma19},
spectra of molecular aggregates~\cite{ren2018time, baiardi2019large, wang2023minimizing},
transport properties of organic semiconductors and thermoelectric materials~\cite{li2021general, ge2022computational}
, as well as the simulation of open quantum systems~\cite{chin2013role, borrelli2017simulation, li2020numerical, yang2023time}.

Matrix product operators (MPO) have significantly contributed to the success of MPS~\cite{schollwock2011density, chan2016matrix}.
Similar to MPS, MPO is a low-rank decomposition of quantum operators.
When the input operators are in a sum-of-product (SOP) form, the analytical form of the MPO can be obtained without any approximation.
The form of MPO for a given operator is not unique, and it is desirable to construct the most compact MPO to minimize the subsequent computational cost in DMRG algorithms.
In the following, we refer to the most compact MPO as the ``optimal'' MPO.
Typically, the construction of the MPO for the Hamiltonian is required for the ground state search or time evolution. 
The MPO of interested physical observables are also frequently constructed for efficient evaluation of the expectation values.
One of the earliest and most straightforward methods for constructing MPOs is through manual design. 
This approach is commonly used in practical MPS algorithms, particularly for constructing MPOs for \textit{ab initio} electronic structure Hamiltonians~\cite{keller2015efficient, chan2016matrix}. 
However, this method is labor-intensive and prone to errors if the MPOs of many different types of operators are required.
To alleviate the need for manual MPO design for different operators, various methods for automated MPO construction based on the input operator have been proposed.
A popular approach involves naively constructing an MPO that is far from optimal initially,
and then compressing it through singular value decomposition (SVD)~\cite{stoudenmire2017sliced, hubig2017generic}.
Recently, another method based on bipartite graph theory has been proposed by us~\cite{ren2020general}.
The method translates from the symbolic operator strings to the optimal MPO without any numerical error
and has become the cornerstone of a series of recent TD-DMRG applications~\cite{li2021general, ge2022computational, wang2023minimizing, jiang2023unified}.

The concept of MPS/MPO can be generalized to TTNS/TTNO~\cite{szalay2015tensor}.
The manual construction of TTNO is more complex than MPO due to the greater flexibility of the TTN structure~\cite{ke2023tree}.
Developing a general algorithm for automatic TTNO construction also presents a significant challenge.
Analog to the construction of MPO, numerical SVD compression can be used for TTNO construction~\cite{otto2014multi, sulz2024numerical}.
However, this numerical SVD compression is typically expensive.
Recently, an algorithm for the symbolic construction of TTNO based on state diagrams is proposed~\cite{milbradt2024state},
yet the resulting TTNO is not optimal.
Due to these challenges, several TTNS studies have opted to use a direct sum-of-product Hamiltonian instead of TTNO~\cite{manthe2008multilayer, murg2010simulating, larsson2019computing, yan2021efficient}, 
which results in a higher computational scaling than using optimal TTNO~\cite{ren2022time}.

In this work, we extend our former bipartite graph theory approach for MPO construction to TTNO construction.
Our algorithm efficiently constructs the optimal TTNO for any operator in the sum-of-product form.
For the spin-boson model~\cite{leggett1987dynamics}, our algorithm generates a TTNO whose $\Mo$ is constant, meaning it's independent of the number of modes in the model.
Combined with the projector splitting method for time evolution, we show that the computational cost for the simulation scales linearly with the number of modes.
We also analyze the computational scaling with respect to the dimension of node indices in TTNS for different TTN topologies, ranging from MPS to binary and ternary trees.
We showcase  the capabilities of our algorithm by studying the spin relaxation dynamics of the spin-boson model and charge transport in a molecular junction.
We also consider the finite temperature effect through thermo field dynamics. 
Thanks to the automatic TTNO construction, the programming cost for the inclusion of the temperature effect is negligible.

\section{Algorithm Implementation}
In this section, we will first recap the concepts of TTNS and TTNO in Sec.~\ref{sec:algo-ttns-ttno}, 
and then describe the time evolution algorithm based on the projector splitting integrator and TTNO in Sec~\ref{sec:algo-ps}.
Finally, we will describe our algorithm for automatic TTNO construction based on bipartite graph theory in Sec~\ref{sec:algo-ttno}.
All of the algorithms discussed in this section have been implemented in the latest version of the open-source package \textsc{Renormalizer}.
The core advantage of our implementation,
compared with other TTNS packages such as the Heidelberg MCTDH package~\cite{mctdh:MLpackage} and QuTree~\cite{ellerbrock2024qutree},
is its use of TTNO and its Python-based nature. Python is a high-level scripting language known for its readability and ease of use. 
This combination of Python's accessibility and TTNO's efficiency makes our implementation a powerful platform for TTNS-based simulations.

\subsection{TTNS and TTNO}
\label{sec:algo-ttns-ttno}
\begin{figure}[htpb]
    \centering    \includegraphics[width=1\textwidth]{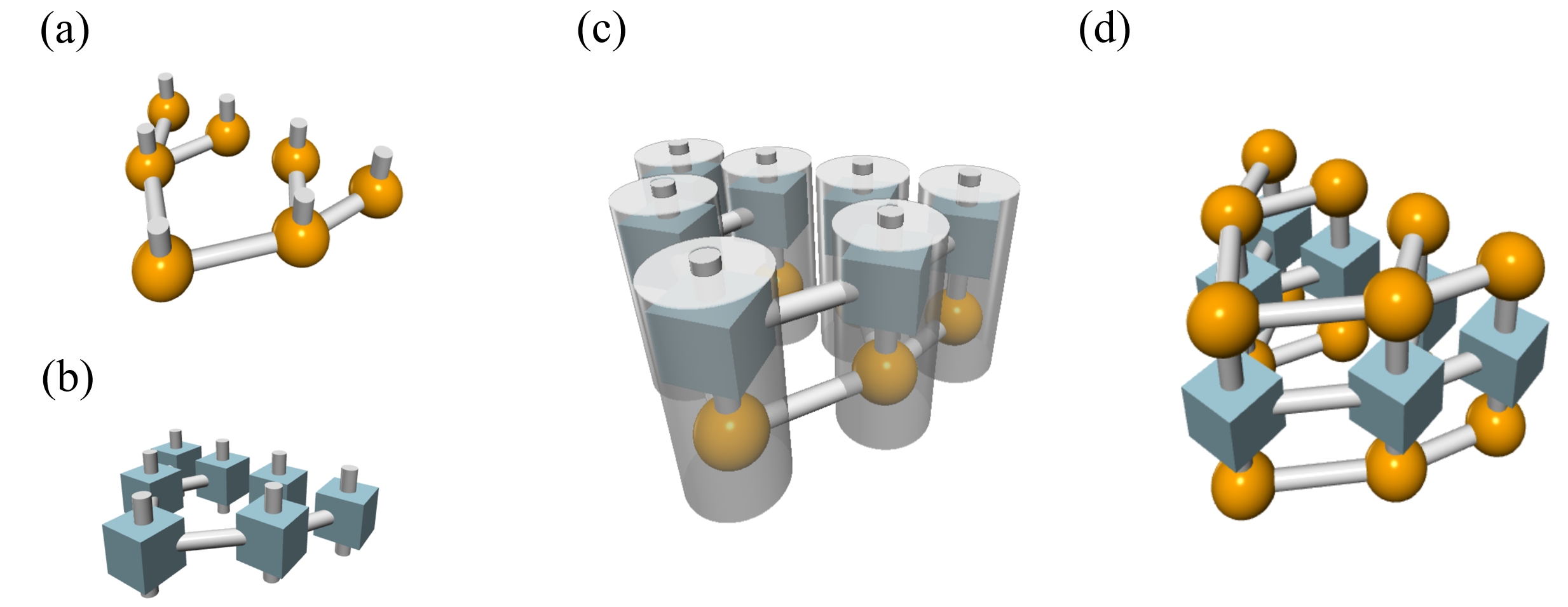}
    \caption{Schematic diagram of TTNS and TTNO using a perfect binary tree as an example. (a) and (b) show the schematic diagram for TTNS and TTNO respectively. Each ball represents a node in TTNS and each box represents a node in TTNO. Vertical cylinders represent physical indices. (c) The tensor network contraction diagram for applying a quantum operator represented by TTNO to a quantum state represented by TTNS. Tensors in grey cylinders are contracted together to form a new TTNS. (d) The tensor network contraction diagram when calculating $\braket{\Psi|\hat O |\Psi}$,  the expectation value of a physical observable.}
    \label{fig:diagram}
\end{figure}

Suppose a quantum system has $N$ degrees of freedom, and for each degree of freedom the corresponding 
primitive basis is $\ket{\sigma_i}$,
the TTNS ansatz represents the wavefunction of this many-body system as the result of contracting low-rank tensors.
The TTNS ansatz can be expressed as:
\begin{equation}
\label{eq:ttns}
    \ket{\Psi} = \sum_{\{a\}, \{\sigma\}} A[1]^{\sigma_1}_{\Lambda_1, a_1} 
    A[2]^{\sigma_2}_{\Lambda_2, a_2} 
    \cdots
    A[N]^{\sigma_N}_{\Lambda_N, a_N} 
    \ket{\sigma_1 \sigma_2 \cdots \sigma_N} \ .
\end{equation}
Here $A[i]$ represents the low-rank tensors, with indices $\sigma_i$, $a_i$ and $\Lambda_i$.
$\sigma_i$ is called the physical index since it is associated with a physical degree of freedom,
whereas $a_i$ and $\Lambda_i$ are virtual indices.
The contraction between $A[i]$ is performed according to a tree topology.
The index $\Lambda_i$ is a collective index that determines the connection topology of the TTNS structure.
$\Lambda_i$ connects to child nodes and $a_i$ connects to the parent node. 
For instance, $\Lambda_i=\{a_{i-1}\}$ means a MPS.
For a perfect binary tree with $2^M-1$ tree nodes, $\Lambda_i=\{a_{2i-1}, a_{2i}\}$ for $i \le 2^{M-1}$ and $\Lambda_i=\varnothing $ otherwise.
A schematic diagram for TTNS is shown in Fig.~\ref{fig:diagram}(a).
In this paper, we denote the dimension of $a_i$ in a TTNS as $\Ms$, and the dimension of $\sigma_i$ as $d$.
$\Ms$ and $d$ are called virtual and physical bond dimension in the MPS language respectively,
and $\Ms$ represents the number of single-particle functions in the MCTDH language.
Larsson recently provided a thorough description for the two sets of languages~\cite{larsson2024tensor}.
In principle, the dimension of $a_i$ and $\sigma_i$ can be different for different nodes in the tree.
For simplicity, we assume the indices for different nodes share the same dimension unless otherwise specified.
The size of $A[i]$ is thus $\Ms^kd$, where $k$ is the number of nodes connected to the $i$th node.

In Eq.~\ref{eq:ttns} we have assumed that each node in the tree is associated with a physical degree of freedom.
However, TTNS can also include entirely ``virtual'' nodes as seen in ML-MCTDH or three-legged tree tensor network states~\cite{gunst2018t3ns}, which is not associated with any physical degree of freedom.
For these virtual nodes, we can assign an auxiliary physical degree of freedom. 
This auxiliary degree of freedom has a Hilbert space of dimension 1, and the only permissible operator is the identity operator. By adopting this approach, the need for special treatment of these nodes is eliminated, and both the notation and the implementation are simplified.

Just as TTNS, a quantum operator $\hat O$ can be expressed as a TTNO
\begin{equation}
\label{eq:ttno}
    \hat O = \sum_{\{a\}, \{\sigma\}, \{\sigma'\}} W[1]^{\sigma'_1,\sigma_1}_{\Lambda_1, a_1} 
    \cdots
    W[N]^{\sigma'_N, \sigma_N}_{\Lambda_N, a_N} 
    \ket{\sigma'_1 \cdots \sigma'_N} \bra{\sigma_1 \cdots \sigma_N} \ .
\end{equation}
Each tensor $W[i]$ in Eq.~\ref{eq:ttno} is a numeric tensor, expanded in the basis of $\ket{\sigma_i}$.
The dimension of $a_i$ in a TTNO is denoted as $\Mo$, and the size of $W[i]$ is $\Mo^k d^2$.
TTNO can also be expressed in \rev{a symbolic} form
\begin{equation}
\label{eq:ttno-sym}
    \hat O = \sum_{\{a\}} \hat W[1]_{\Lambda_1, a_1} 
    \cdots
    \hat W[N]_{\Lambda_N, a_N} \ ,
\end{equation}
where $\hat W[i]$ is a tensor whose elements are symbolic operators.
The size of $\hat W[i]$ is $\Mo^k$.
A schematic diagram for TTNO is shown in Fig.~\ref{fig:diagram}(b).
For later convenience, we define $\hat W[\sim i]$ based on the recurring relation
\begin{equation}
\label{eq:w-recur}
    \hat W[\sim i]_{a_i} = \sum_{\Lambda_i} \hat W[i]_{\Lambda_i, a_i} \prod_{j\in\textrm{child}(i)} \hat W[\sim j]_{a_j} \ .
\end{equation}
Here $\textrm{child}(i)$ refers to the indices for all direct children of the $i$th tree node,
and $\Lambda_i = \{a_j|j\in\textrm{child}(i)\}$.
Note that for leaf nodes we have $\hat W[\sim i]_{a_i} = \hat W[i]_{\varnothing, a_i} $.

Similar to MPO/MPS, a TTNO can be applied to a TTNS through tensor contractions, resulting in a new TTNS,
as shown in Fig.~\ref{fig:diagram}(c).
This feature provides significant flexibility for manipulating TTNS 
and lays the foundation of a whole class of time evolution methods based on propagation and compression~\cite{garcia2006time, paeckel2019time, ren2022time}.
Another immediate advantage of using TTNO is the efficient computation of the physical observable $\braket{\Psi|\hat O|\Psi}$ through tensor network contraction.
To achieve this, we stack $\ket{\Psi}$, $\ket{\hat O}$ and $\bra{\Psi}$ in a three-layer manner, as shown in Fig.~\ref{fig:diagram}(d).
The contraction process then begins from the
leaves and moves inward towards the root.
The computation cost scales polynomially with $\Ms$, $\Mo$ and $d$.

\subsection{Time Evolution based on Projector Splitting}
\label{sec:algo-ps}
\begin{figure}[htpb]
    \centering    \includegraphics[width=\textwidth]{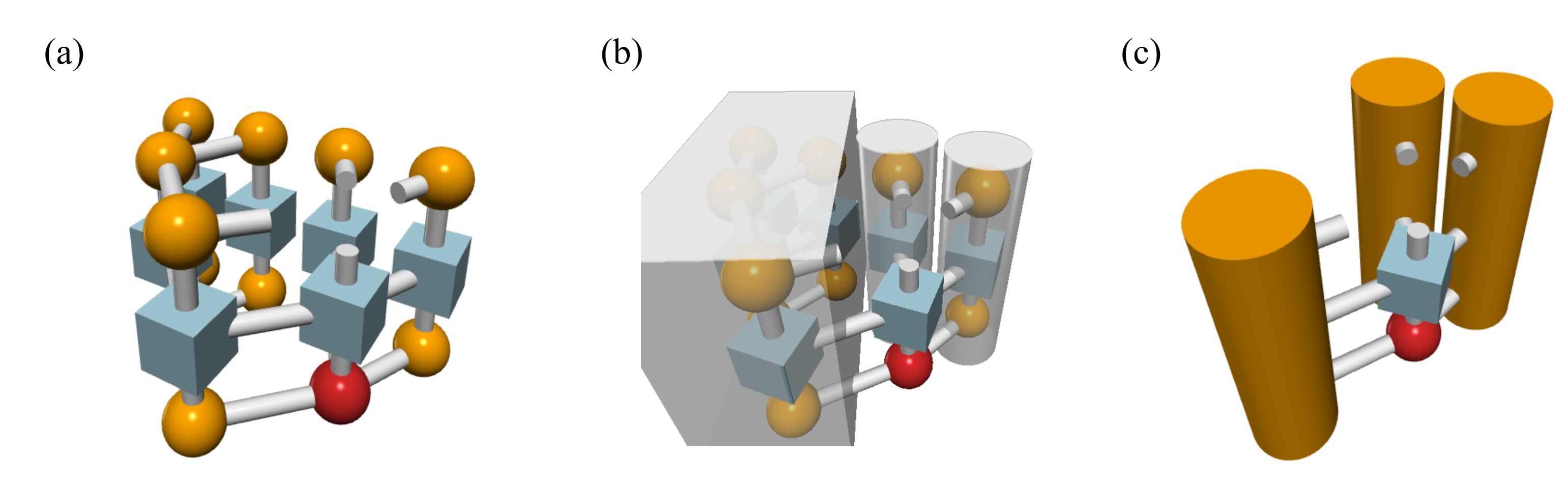}
    \caption{Schematic diagram for the calculation of $\hat H^{\textrm{eff}}_i A[i]$ using TTNS and TTNO. 
    (a) The red ball represents $A[i]$ and the rest of the diagram constitutes $\hat H^{\textrm{eff}}_i$.
    (b) Tensors in grey containers are contracted to construct the effective environment in $\hat H^{\textrm{eff}}_i$.
    (c) The actual tensors involved in the computation of $\hat H^{\textrm{eff}}_i A[i]$ and orange cylinders are environment tensors.}
    \label{fig:diagram2}
\end{figure}

Based on time-dependent variational principle~\cite{broeckhove1988equivalence}, 
the formal solution of the time-dependent Schr\"odinger equation
for tensor networks can be expressed as~\cite{haegeman2016unifying}
\begin{equation}
\pdv{\ket{\Psi}}{t} = -i\hat P \hat H \ket{\Psi} \ .
\end{equation}
Here $\hat P=\sum_i^N\hat P^+_i-\sum_i^{N-1} \hat P^-_i$ is a projection operator to the tangent space of the TTNS manifold.
The exact form of $\hat P$ is provided in the literature~\cite{bauernfeind2020time, larsson2024tensor}.
At a short time step $\dt$, the formal solution of the Schr\"odinger equation is
\begin{equation}
    \ket{\Psi}(t_0+\dt) = e^{-i\hat P \hat H \dt}\ket{\Psi}(t_0) \ .
\end{equation}
The projector splitting integrator decomposes the formal time evolution operator into the product of operators
\begin{equation}
    e^{-i\hat P \hat H \dt} = \prod_i^{N-1} \{e^{-i\hat P^+_i \hat H \dt} e^{i\hat P^-_i \hat H \dt} \} e^{-i\hat P^+_N \hat H \dt} + \mathcal{O}(\dt^2) \ .
\end{equation}
Each term in the product corresponds to the time evolution of a local tensor in the TTNS.
The complete time evolution consists of a sweep over the tree tensor network.
For example, $e^{-i\hat P^+_i \hat H \dt}$ is implemented as a local time evolution of 
$A[i]$ based on the effective Hamiltonian $\hat H^{\textrm{eff}}_i$
\begin{equation}
    A[i](t_0+\dt) = e^{-i \hat H^{\textrm{eff}}_i \dt}A[i](t_0) \ .
\end{equation}
A schematic illustration of $\hat H^{\textrm{eff}}_i A[i]$ is shown in Fig.~\ref{fig:diagram2}(a).
In Fig.~\ref{fig:diagram2},
$A[i]$ is shown  as the red ball and $\hat H^{\textrm{eff}}_i$ corresponds to the rest of the diagram.
The term $e^{-i \hat H^{\textrm{eff}}_i \dt}A[i]$ can then be computed using a Krylov matrix exponential solver.

The computational bottleneck for the time evolution is the contraction between $A[i]$ and $\hat H^{\textrm{eff}}_i$.
During the contraction, the tensors in the grey containers in Fig.~\ref{fig:diagram2}(b) are first contracted together,
resulting in the contraction pattern shown in Fig.~\ref{fig:diagram2}(c).
The construction of the environment tensor is not the bottleneck of the computation. 
During the sweep process, the environment tensors from the previous step can be employed to calculate the environment tensors required for the next step.
The computation of the environment tensor has lower scaling than computing $\hat H^{\textrm{eff}}_i A[i]$,
which involves the contraction between $k$ environment tensors, $W[i]$ and $A[i]$.
The size of the environment tensor is $\Ms \times \Mo \times \Ms=\Ms^2 \Mo$.
Recall that the size of $A[i]$ is $\Ms^k d$ and the size of $W[i]$ is $\Mo^kd^2$.
The contraction between one of the environment tensor and $A[i]$ has computational cost $\mathcal{O}(\Ms^{k+1}\Mo d)$
and the resulting tensor is of size $\Ms^k\Mo d$.
The tensor is subsequently contracted with $W[i]$ at a computational cost of $\mathcal{O}(\Ms^{k}\Mo^k d^2)$.
The resulting tensor of size $\Ms^k\Mo^{k-1} d$ is then contracted with another environment tensor
at a computational cost of $\mathcal{O}(\Ms^{k+1} \Mo^{k-1} d)$.
The resulting tensor has a smaller size of $\Ms^k\Mo^{k-2} d$ and the rest of the contraction 
with other environment tensors has less computational cost.
In all cases discussed in this work, $\Mo$ is less than 10, and $\Ms$ ranges from 20 to 96.
Meanwhile, most of $d$ is less than or equals to 10.
Thus we can assume $\Mo \approx d \ll \Ms$ and the overall computational scaling for the contraction of
$\hat H^{\textrm{eff}}_i A[i]$ is $\mathcal{O}(\Ms^{k+1} \Mo^{k-1} d)$.
The complete time evolution also requires the application of $e^{-i\hat P^-_i \hat H \dt}$, 
which is similar to the application of $e^{i\hat P^+_i \hat H \dt}$ but with a reduced cost. 
In our implementation, we use a second-order symmetric Trotter decomposition of $e^{-i\hat P \hat H \dt}$. 
For the complete algorithm of the time evolution, please refer to the cited papers~\cite{bauernfeind2020time, lindoy2021time, lindoy2021time2}.

\subsection{Automatic Construction of TTNO}
\label{sec:algo-ttno}
In this section, we describe our algorithm for the automatic construction of TTNO based on bipartite graph theory.
Our algorithm is developed based on the algorithm for the automatic construction of MPO~\cite{ren2020general}.
Any quantum operator in the SOP form can be expressed as
\begin{equation}
 \label{eq:sop} 
    \hat{O}  = \sum_{o=1}^K \gamma_{o} \prod_{i=1}^N \hat{z}_i^{(o)} , 
\end{equation}
where $K$ is the number of terms, $N$ is the number of degrees of freedom, $\gamma_{o}$ stands for the coefficient for the $o$th term and $\hat{z}_i^{(o)}$ is the elementary operator for the $i$th degree of freedom for the $o$th term.
Note that $\hat{z}_i^{(o)}$ can be an identity operator.
Tensor networks usually impose the commutation condition $[\hat z_i, \hat z_j] = 0$ for $i \neq j$.
In Sec.~\ref{sec:transport}, we discuss how to comply with the anti-commutation property for fermion operators.
Table~\ref{tab:sop} shows the tabular form of $\hat O$,
where each row represents a term in $\hat O$.
We refer to Table~\ref{tab:sop} as the SOP table in the following discussion.
The coefficients $\gamma_0$ can be stored in another table, which has the same number of rows as Table~\ref{tab:sop} but has only one column.
\begin{table}[hbpt]
\caption{\rev{Tabular} form of the SOP operator Eq.~\ref{eq:sop}. Each row represents a term in the operator and each column corresponds to one of the degrees of freedom.}
\label{tab:sop}
\begin{tabular}{cccccc}
\hline
 &  \multicolumn{5}{c}{Degree of freedom} \\
Term index & 1 & ... & $i$ & ... & $N$ \\
\hline
 1 & $\hat z_1^{(1)}$ & ... & $\hat z_i^{(1)}$ & ... & $\hat z_N^{(1)}$ \\
 ... & ... & ... & ... & ... & ... \\
 $o$ & $\hat z_1^{(o)}$ & ... & $\hat z_i^{(o)}$ & ... & $\hat z_N^{(o)}$ \\
 ... & ... & ... & ... & ... & ... \\
 $K$ & $\hat z_1^{(K)}$ & ... & $\hat z_i^{(K)}$ & ... & $\hat z_N^{(K)}$ \\
\hline
\end{tabular}
\end{table}


In tree tensor networks, cutting an arbitrary edge will divide the tree into two parts.
Suppose by cutting the $i$th edge, which connects the $i$th node to its parent,
the degrees of freedom of the entire system are divided into two parts, namely $L$ and $R$.
It is then possible to express $\hat O$ as
\begin{equation}
     \hat{O} = \sum_{o_i=1}^K  \gamma_{o_i} \cdot \hat{O}^{(o_i)}_{i, L} \otimes \hat{O}^{(o_i)}_{i,R} , \label{eq:split}    
\end{equation} 
where
\begin{equation}
\begin{aligned}
    \hat{O}^{(o_i)}_{i, L}  &= \prod_{j \in \Omega^{i, L}} \hat{z}_j^{(o_i)} ,\\
    \hat{O}^{(o_i)}_{i,R} & = \prod_{j\in \Omega^{i, R}} \hat{z}_j^{(o_i)}.
\end{aligned}
\end{equation}
Here,  $\Omega^{i, L(R)}$ stands for the set of degrees of freedom that fall into the $L(R)$ part after cutting the $i$th edge.
Based on Eq.~\ref{eq:split}, the $\Mo$ for the TTNO at the $i$th edge is $K$, which showcases a \rev{straightforward} construction of TTNO that is not optimal.

The key to the construction of optimal TTNO is to exploit redundancy in $\{\hat{O}^{(o_i)}_{i, L(R)}|o_i\in[1,K]\}$.
Denote unique operators in $\{\hat{O}^{(o_i)}_{i, L(R)}|o_i\in[1,K]\}$ as $\mathcal{U}^{i, L(R)}$,
$\hat O$ can be written as
\begin{equation}
    \hat{O} = \sum_{jk}\Gamma_{jk} \mathcal{U}^{i, L}_j \otimes \mathcal{U}^{i, R}_k \ ,
\end{equation}
where $\Gamma$ is the coefficient matrix on the basis of $ \mathcal{U}^{i, L} \otimes \mathcal{U}^{i, R}$.
A natural next step is to perform Schmidt decomposition to reduce the dimension and obtain a more efficient TTNO,
similar to the idea of TTNS.
The computational scaling of this approach is $\mathcal{O}(K^3)$ since $\Gamma$ is approximately of size $K\times K$.
The $\Mo$ at the $i$th edge equals the rank of $\Gamma$.
Additional truncation based on singular values can be performed to compress the TTNO~\cite{zhai2023block2}.

In addition to the Schmidt decomposition method, better scaling can be obtained by leveraging the sparse nature of $\Gamma$ using the bipartite graph theory. $\Gamma$ is sparse since it only contains $K$ non-zero elements.
A bipartite graph is a type of graph where all vertices can be divided into two distinct sets, and every edge connects a vertex in 
one set to a vertex in the other set.
In this context,
$\mathcal{U}^{i, L} $ and $ \mathcal{U}^{i, R}$ are considered as the two sets of vertices in the bipartite graph, and their interactions represented by nonzero elements in $\Gamma$ are considered as the graph edges.
Our goal is to find the minimum number of vertices that can connect to all edges in the graph,
a problem known as the minimum vertex cover problem in graph theory.
The minimum $\Mo$ at the $i$th edge equals the number of these vertices, by constructing complementary operators~\cite{xiang1996density} for the vertices.
This minimum number of vertices can be efficiently found in polynomial time with respect to the number of vertices and edges.
Specifically, since there are $K$ vertices and $K$ edges, the computational scaling is $\mathcal{O}(K^{\frac{3}{2}})$  using the Hopcroft-Karp algorithm~\cite{Hopcroft73}.
This is an improvement over the $K^3$ scaling using SVD.
Solving the minimum vertex cover problem for the $i$th edge determines the construction of optimal TTNO for the $i$th node by
constructing complementary operators according to the solution of the minimum vertex cover problem.
The entire TTNO is constructed by iterating over all of the nodes in the tree.
For the simulation of open quantum systems such as the spin-boson model and the transport model studied in this paper,
the number of terms in the Hamiltonian $K$ is proportional to the number of nodes and edges in the tree tensor network $N$.
Therefore, the overall scaling for the construction of the complete TTNO is $N^{\frac{3}{2}}\times N = N^{\frac{5}{2}}$.

The construction of MPO/TTNO based on the bipartite graph approach does not rely on the coefficients of the SOP operator.
In other words, with different but finite $\gamma_o$, the MPO/TTNO has the same $\Mo$ for the same operator.
The approach of manual MPO/TTNO design also shares this feature.
However, the same is not true for the SVD compression approach.
If $\gamma_o$ exhibits a particular pattern, it is possible to obtain a more compact MPO/TTNO based on SVD compression without numerical compression error.
Consider, for example, that $\Gamma$ is a $2\times 2$ matrix and all matrix elements are 1.
The minimum number of vertices to connect to all edges is thus 2.
Meanwhile, SVD over $\Gamma$ yields only one non-zero singular value.
Such a pattern is hardly encountered in chemical models in practice.
\rev{However, for lattice models encountered in physics, most Hamiltonian terms share the same coefficient. In such cases, SVD or QR over $\Gamma$ is preferred for obtaining the optimal MPO/TTNO}.

To illustrate the concept of the SOP table, bipartite graph, and minimum vertex cover, let's consider a spin-boson model where a spin is coupled with a set of vibrational modes. 
The Hamiltonian for the spin-boson model is given by:
\begin{equation}
\label{eq:ham-sbm}
    \hat H = \Delta \hat \sigma^x  + \frac{1}{2}\sum_{i=1}^{N_b} (\hat p^2_i + \omega_i^2\hat q_i^2) + \hat \sigma^z\sum_{i=1}^{N_b} c_i \hat q_i \ ,
\end{equation}
which follows the sum-of-product form.
Here $\hat p$ and $\hat q$ are the momentum and coordinate operators of the vibration modes.
$\omega$ is the vibration frequency. $c$ is the coupling constant. $\hat \sigma^x$ and $\hat \sigma^z$ are Pauli matrices for the spin. $\Delta$ is the tunneling constant.
For ease of demonstration, we limit $N_b=2$, and generalization to more complex cases is straightforward.
The two vibration modes are denoted as $v_1$ and $v_2$.
\rev{
We consider a simple tree where the spin node is the root, and the vibration nodes are the two leaves.
}

We can first construct the SOP table for Eq.~\ref{eq:ham-sbm} as shown in Table~\ref{tab:sbm-sop}.
Next, assume spin and $v_1$ are in the $L$ part while $v_2$ is in the $R$ part, 
as indicated by the vertical line in Table~\ref{tab:sbm-sop}.
The operators in $\Omega^L$ and $\Omega^R$ are redundant. 
For example, the
unique operators in $\Omega^R$, denoted as $\mathcal{U}^{R}$, are $\{\hat I, \hat p^2_2, \hat q^2_2, \hat q_2\}$.
Meanwhile, $\Omega^L$ has 6 unique operators.
$\Gamma$ is then a $6\times 4$ matrix with 7 nonzero elements.
SVD over $\Gamma$ yields 3 nonzero singular values, which means the minimum $\Mo$ of TTNO is 3.

\begin{table}[hbpt]
\caption{SOP table of the Hamiltonian of the spin-boson model Eq.~\ref{eq:ham-sbm} with $N_b=2$, as a special case of Table~\ref{tab:sop}. }
\label{tab:sbm-sop}
\begin{tabular}{|c|cc|c|}
\hline
 & spin & $v_1$ & $v_2$\\ 
\hline
 1 & $\hat \sigma^x$ & $\hat I$ & $\hat I$ \\
 2 & $\hat I$ & $\hat p^2_1$ & $\hat I  $ \\
 3 & $\hat I$ & $\hat q^2_1$ & $\hat I  $ \\ 
 4 & $\hat I$ & $\hat I  $ & $\hat p^2_2$ \\
 5 & $\hat I$ & $\hat I  $ & $\hat q^2_2$ \\ 
 6 & $\hat \sigma^z$ & $\hat q_1 $ & $\hat I$ \\
 7 & $\hat \sigma^z$ & $\hat I $ & $\hat q_2$ \\
\hline
\end{tabular}
\end{table}

\begin{figure}[hbpt]
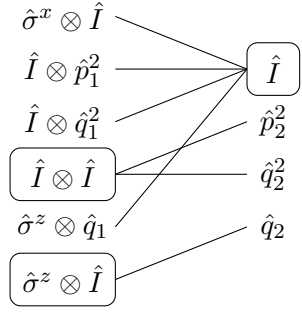

  \centering
    \begin{diagram}

    \draw (0, 5) node {$\hat \sigma^x \otimes \hat I$};
    \draw (0, 4) node {$\hat I \otimes \hat p^2_1$};
    \draw (0, 3) node {$\hat I \otimes \hat q^2_1$};
    \draw (0, 2) node {$\hat I \otimes \hat I$};
    \draw[rounded corners] (0-1, 2+0.5) rectangle (0+1, 2-0.5);
    \draw (0, 1) node {$\hat \sigma^z \otimes \hat q_1$};
    \draw (0, 0) node (X) {$\hat \sigma^z \otimes \hat I$};
    \draw[rounded corners] (0-1, 0+0.5) rectangle (0+1, 0-0.5);

    \draw (1,5) -- (3.5,4);
    \draw (1,4) -- (3.5,4);
    \draw (1,3) -- (3.5,4);
    \draw (1,1) -- (3.5,4);

    \draw (1,2) -- (3.5,3);    
    \draw (1,2) -- (3.5,2);    

    \draw (1,0) -- (3.5,1);    

    \draw (4, 4) node {$\hat I$};
    \draw[rounded corners] (4-0.5, 4+0.5) rectangle (4+0.5, 4-0.5);
    \draw (4, 3) node {$\hat p^2_2$};
    \draw (4, 2) node {$\hat q^2_2$};
    \draw (4, 1) node {$\hat q_2$};

    \end{diagram}  
    
  \caption{The bipartite graph for SOP operators taking the Hamiltonian of the spin-boson model with $N_b=2$ as an example.
  The operators acting on the spin and $v_1$ are grouped on the left and the operators acting on $v_2$ are grouped on the right.
  Each edge represents a term in the Hamiltonian and the operators in boxes are a minimal set of vertices that covers all edges.}
  \label{fig:sbm-bipartite}
\end{figure}

Fig.~\ref{fig:sbm-bipartite} shows the bipartite graph representation of Table~\ref{tab:sbm-sop}.
The vertices on the left(right) side are $\mathcal{U}^{L(R)}$, and
 each edge represents a term in Table~\ref{tab:sbm-sop}.
For optimal TTNO, it is necessary to select a minimum set of  vertices in the graph that covers all edges.
Such a task can be solved efficiently using the Hopcroft-Karp algorithm.
By selecting $\hat I \otimes \hat I$, $\hat \sigma^z \otimes \hat I$ in the $L$ set and $\hat I$ in the $R$ set,
i.e. operators in round-corner boxes,
all edges are covered.
Complementary operators are then constructed based on the edge connections.
For example, since $\hat I \otimes \hat I$ is connected to $\hat p^2_2$ and $\hat q^2_2$, these two are combined to form a complementary operator $\frac{1}{2}\hat p^2_2 + \frac{1}{2}\omega_2^2 \hat q^2_2$ for $\hat I \otimes \hat I$.
\rev{
This allows us to determine the TTNO tensor $\hat W[v_2]$ for the node representing $v_2$.
The tensor has only one virtual index and its symbolic form is
\begin{equation}
    \hat W[v_2] = \mqty[\hat I & \frac{1}{2}\hat p^2_2 + \frac{1}{2}\omega_2^2 \hat q^2_2 & c_2 \hat q_2]
\end{equation}

In order to determine the whole TTNO, we additionally bipartite the system by cutting the edge between the spin node and the $v_1$ node. Then the left part of the system contains $v_1$
and the right part contains the spin and $v_2$.
Similarly, $\hat W[v_1]$ is determined as
\begin{equation}
    \hat W[v_1] = \mqty[\hat I & \frac{1}{2}\hat p^2_1 + \frac{1}{2}\omega_1^2 \hat q^2_1 & c_1 \hat q_1]
\end{equation}

Since the spin node is the root node, its symbolic TTNO tensor $\hat W[\textrm{spin}]$ can be determined from
$\hat W[\sim v_1] = \hat W[v_1]$, $\hat W[\sim v_2] = \hat W[v_2]$ and the overall SOP Hamiltonian $\hat H$. $\hat W[\textrm{spin}]$ is then 
\begin{equation}
    \hat W[\textrm{spin}] = \mqty[\Delta \hat \sigma^x & \hat I & \hat \sigma^z \\
    \hat I & 0 & 0 \\
    \hat \sigma^z& 0 & 0 ]
\end{equation}
}

In the specific case of spin-boson model, it can be shown that $\Mo=3$ regardless of the number of discretized modes.
Consequently, for fixed $\Ms$ and $d$, the computational cost for updating one of the tensors $A[i]$ is independent of the number of modes.
Therefore, the total computational time for a single time evolution step scales linearly with the number of nodes in the tree or the number of modes in the system.

Finally, we outline the general procedure to construct TTNO using the bipartite graph theory step by step.
The algorithm visits all nodes in the post-order,
meaning that if the $i$th node is visited, all of its children have already been visited
and $\hat W[\sim j]$ for all of the children have been constructed.
For the $i$th node in the tree in the post-order, perform the following steps:
\begin{enumerate}
    \item Construct the SOP table of the target operator $\hat O$.
    Consider elementary operators from all children, as provided by $\hat W[\sim j]$, at the current node, as well as the rest of the system.
    If the current node does not have children, identity operators can be taken as placeholders for the elementary operators from the children.
    \item Divide the SOP table into two parts. The left part contains elementary operators from all children $\hat z^{(o)}_{\textrm{child}}$ and the current node $\hat z^{(o)}_{\textrm{node}}$. The right part contains all operators from the rest of the system $\hat z^{(o)}_{\textrm{others}}$. Identify the unique operators $\mathcal{U}^{L(R)}$ and use them to
    construct the bipartite graph. 
    Then, solve the minimum vertex cover problem using the Hopcroft-Karp algorithm.
    \item For each operator in $\mathcal{U}^{L}$, if $\mathcal{U}^{L}_k$ is included in the solution of the minimum vertex cover problem,
    append $\mathcal{U}^{L}_k$  as a new row of $\hat W[\sim i]$. 
    For each operator in $\mathcal{U}^{R}$, if $\mathcal{U}^{R}_k$ is included in the solution of the minimum vertex cover problem, add the operators in $\mathcal{U}^{L}$ that are connected to $\mathcal{U}^{R}_k$
    and append the summed operator as a new row of $\hat W[\sim i]$.
    Note that if an edge is connected to both $\mathcal{U}^L$ and $\mathcal{U}^R$, it should only be included once.
    \item Using $\hat W[\sim i]$ and $\hat W[\sim j]$ for all child node of the $i$th node, construct $\hat W[i]$ based on the recurring relation Eq.~\ref{eq:w-recur}. $\hat W[\sim i]$ is then considered as elementary operators for the parent of the $i$th node.
    \item Proceed to the next node following the post-order sequence.
\end{enumerate}

\section{Numerical Results}
In this section, we present numerical results based on our TTNS algorithm with optimal TTNO.
First, we'll demonstrate numerically the computational scaling of our algorithm, highlighting how the computational cost scales linearly with the number of degrees of freedom in the spin-boson model. 
Subsequently, we'll showcase two example simulations of open quantum systems to illustrate the versatility and efficiency of our algorithm.

\subsection{Computational Scaling}
\label{sec:scaling}
We first show the computational scaling of our algorithm through numerical simulation for the spin relaxation dynamics of the spin-boson model.
The Hamiltonian of the model is presented in Eq.~\ref{eq:ham-sbm}.
The coupling between  the spin and the vibrational environment is specified by the spectral density function
\begin{equation}
\label{eq:sdf}
    J(\omega) = \frac{\pi}{2} \sum_j \frac{c_j^2}{\omega_j} \delta(\omega-\omega_j) \ .
\end{equation}
In this section, we focus on the sub-Ohmic spectral density function
\begin{equation}
\label{eq:sdf-ohmic}
    J_{\textrm{Ohmic}}(\omega) = \frac{\pi}{2} \alpha \omega^s\omega_c^{1-s}e^{-\omega/\omega_c}.
\end{equation}
Here $\alpha$ is the dimensionless Kondo parameter that controls the strength of the system-bath coupling.
$\omega_c$ is the characteristic frequency of the bath.
$s$ controls the shape of the spectral density.
A value of $0<s<1$ corresponds to sub-Ohmic spectral density, $s=1$ to Ohmic spectral density, and $s>1$ to super-Ohmic spectral density.
For our purposes, we set $s=0.5$ and $\omega_c=20\Delta$.
$\alpha$ is set to 0.05 unless otherwise specified.
The vibration modes are discretized based on a particular density of states $\rho(\omega)$
\begin{equation}
\label{eq:discre}
    \int_0^{\omega_j} \rho(\omega) d\omega = j, \ j=1,\dots, N_b \ ,
\end{equation}
where $N_b$ is the number of bath modes. The density of states $\rho(\omega)$ is defined as
\begin{equation}
    \rho(\omega) = \frac{N_b+1}{\omega_c} e^{-\omega/\omega_c} \ .
\end{equation}
We use  10 harmonic oscillator eigenbasis for all vibration modes ($d=10$).
This setup is chosen for the scaling benchmark 
to facilitate direct comparison with the results from the literature~\cite{wang2008coherent, wang2010coherent, ren2022time}.
Benchmark results with $\alpha$ from 0.05 to 1.0 are shown in Fig.~\ref{fig:subohmic} in the Appendix, aligning perfectly with previous reports.

\begin{figure}[htpb]
    \centering    \includegraphics[width=.8\textwidth]{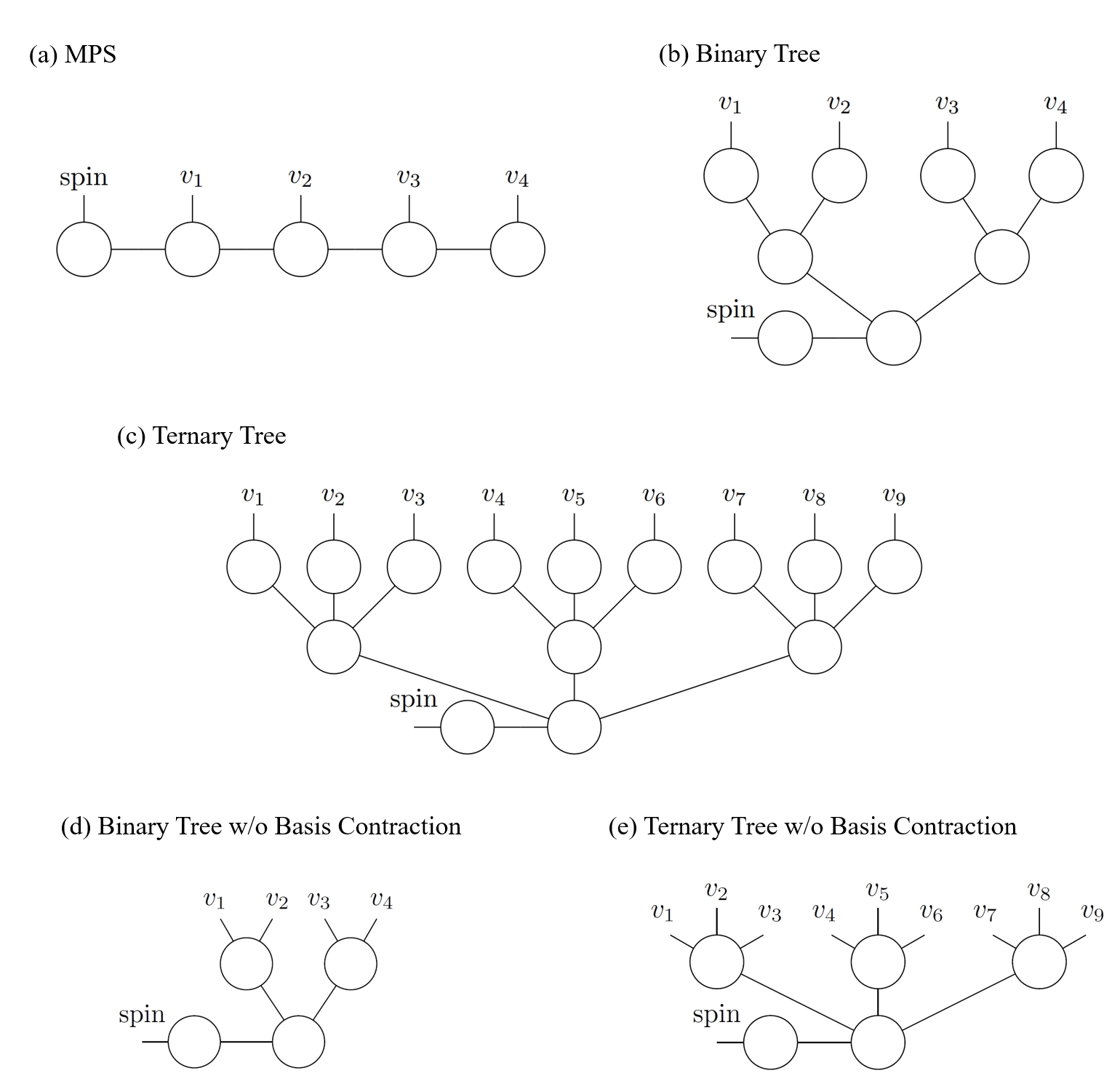}
    \caption{Tree tensor network structures used in this work. (a) linear chain topology, which is identical to MPS. (b) Binary tree topology. (c) Ternary tree topology. \rev{ (d) Binary tree topology without primitive basis contraction. (e) Tinary tree topology without primitive basis contraction.}}
    \label{fig:tree_structure}
\end{figure}

We next describe the TTNS tree structure employed for the simulation.
Our algorithm and its implementation can work with any tree structure, regardless of tree depth, and
the tree structure we've used here is chosen for its simplicity.
It is important to note that these structures may not necessarily be the optimal tree structure for the models we're studying. 
We first look at the MPS structure, where all degrees of freedom are arranged in a linear chain. 
The spin is positioned at the start of the chain, followed by the vibration modes in ascending order of vibration frequency. A visual representation of the MPS structure is provided in Fig.~\ref{fig:tree_structure}(a).
We then describe two types of tree topologies that are inspired by ML-MCTDH.
These are depicted in Fig.~\ref{fig:tree_structure}(b) and Fig.~\ref{fig:tree_structure}(c).
In both types of tree the vibration degrees of freedom are first grouped to form either a binary or ternary tree.
The spin is then attached to the root of the tree.
In both trees, only the leaf node has physical indices.
In Fig.~\ref{fig:tree_structure}(b) and Fig.~\ref{fig:tree_structure}(c),
each primitive mode is first contracted to $\Ms$ states before being connected to the next layer.
The actual tree structure employed in our simulation is a bit more subtle.
More specifically, the contraction of the primitive modes in Fig.~\ref{fig:tree_structure}(b) and Fig.~\ref{fig:tree_structure}(c) applies when $\Ms < d$.
When $\Ms > d$, such contraction is unnecessary and two or three primitive modes are directly linked to the leaf node.
In the tree structure, the number of layers is approximately $\log N_b$.
For all structures considered, the number of nodes is on the same order as $N_b$ or the number of degrees of freedom in the model.

\begin{figure}[htpb]
    \centering    \includegraphics[width=.4\textwidth]{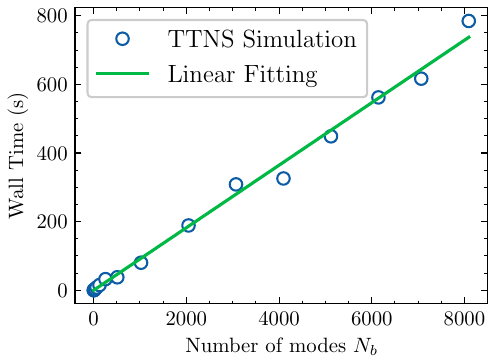}
    \caption{Wall time for one step of the TDVP-PS time evolution based on the spin-boson model as a function of the number of modes $N_b$. $\Ms=20$, $d=10$ and $\dt \Delta = 0.05$. The solid line is a linear fitting of the collected data.}
    \label{fig:wall_time}
\end{figure}

We first show the wall time required for one step of TDVP-PS time evolution versus the number of modes $N_b$ in the model in Fig.~\ref{fig:wall_time}.
We use the binary tree shown in Fig.~\ref{fig:tree_structure}(b), with
$\Ms$ set to 20 and the time evolution step size at $\dt \Delta = 0.05$.
The simulation is run for 10 steps and the average wall time per step is reported.
The computation is carried out on a single core of Intel(R) Xeon(R) Platinum 8255C CPU @ 2.50GHz (the same hereinafter).
Simulations up to $N_b=8096$ reveal that the wall time required to perform a single step of the time evolution scales linearly with the number of bath modes $N_b$ in the model.
This favorable scaling is realized by constructing the Hamiltonian as TTNO with a constant $\Mo$ of 3.
When $N_b=8096$, the average wall time per step is approximately 800 seconds.
Since simulation to $t\Delta=40$ requires 800 steps, a complete simulation with $N_b=8096$ would take roughly one week.
The mode combination technique in ML-MCTDH is not employed in this case.

\begin{figure}[htpb]
    \centering    \includegraphics[width=\textwidth]{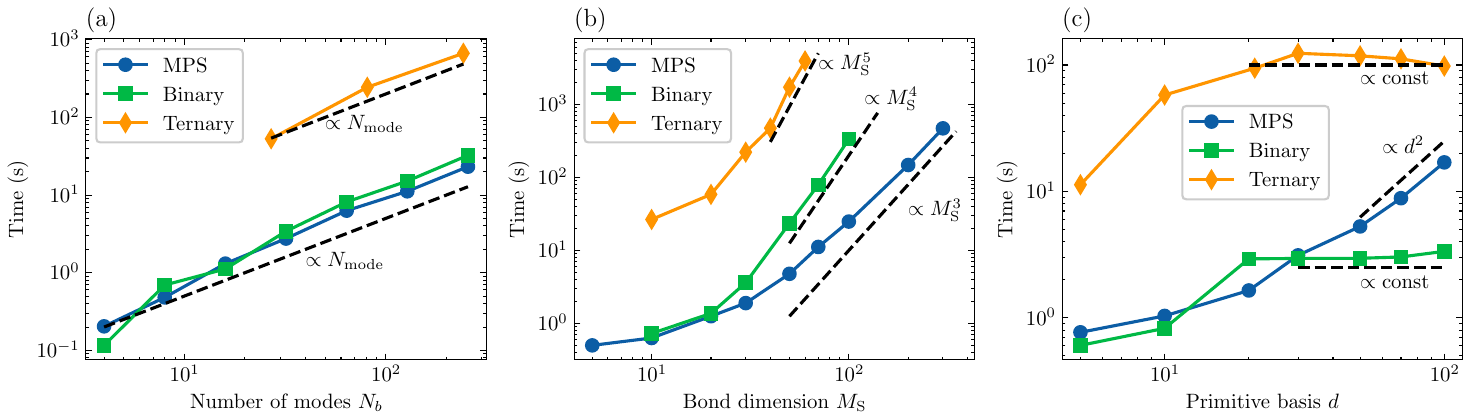}
    \caption{The computational cost of single evolution step using MPS, binary TTNS, and ternary TTNS. All calculations are performed on a single Intel Xeon(R) Platinum 8255C CPU @ 2.50GHz core.  (a) $\Ms=20$, $d=10$ and $N_b$ varies. (b) For MPS and binary TTNS $N_b=16$ and for ternary TTNS $N_b=27$. $d=10$ and $\Ms$ varies. (c) For MPS and binary TTNS $N_b=16$ and for ternary TTNS $N_b=27$. $\Ms=20$ and  $d$ varies. The black dashed lines are guides to indicate the polynomial scaling.}
    \label{fig:scaling_sbm}
\end{figure}

We next examine how different tree structures, as well as the values of $\Ms$ and $d$, affect the computational cost.
In Fig.~\ref{fig:scaling_sbm}(a) we show the computational wall time versus $N_b$ with different tree structures.
The MPS tree structure and the binary tree structure have almost identical computational costs.
This is because, with $\Ms > d$, each leaf node has 2 physical indices in the binary tree.
Thus, both MPS and the binary tree have approximately the same number of nodes.
Additionally, 
in both MPS and binary tree most nodes have three indices (virtual and physical combined). 
This is why the two different tree structures have almost the same computational cost.
In the ternary tree, most nodes have 4 physical and virtual indices, leading to a significantly higher computational cost compared to the other two cases.
In Fig.~\ref{fig:scaling_sbm}(b) we show the computational wall time versus $\Ms$.
As analyzed earlier in Sec.~\ref{sec:algo-ps}, the computational cost scales as $\mathcal{O}(\Ms^{k+1})$ with respect to $\Ms$ where $k$ is the number of nodes connected to each node.
Clearly, for MPS $k=2$, for the binary tree $k=3$, and for the ternary tree $k=4$.
The computational scaling is consistent with our analysis in Fig.~\ref{fig:scaling_sbm}.
In Fig.~\ref{fig:scaling_sbm}(c) we show the computational wall time versus the number of primitive basis $d$.
According to our analysis in Sec.~\ref{sec:algo-ps}, the computational cost scales as $\mathcal{O}(d)$ when $d$ is small, and $\mathcal{O}(d^2)$ when $d$ is large.
However, this analysis assumes that each node in the TTNS has a physical index with dimension $d$,
which is true for MPS but not true for the tree networks employed here.
For binary and ternary trees, when $d$ is large, the $d$ primitive basis is firstly contracted to $\Ms$ states at the leaf node.
Thus, up to reasonably large $d$, the computational cost with respect to these leaf nodes is negligible compared with the computation of the body nodes, which have more indices than the leaf nodes.
\rev{
For the cases considered in this work, TTNO and MPO have similar bond dimensions.  The key advantage of using TTNO is that they can reduce the computing scaling over the number of primitive basis $d$ to a constant. For comparison, the MPS/MPO approach has $\mathcal{O}(d^2)$ scaling. This significant improvement in scaling justifies the development and use of TTNS/TTNO approaches, especially when $d$ is large, despite their increased coding complexity.
This advantage helps explain why MPS is predominantly used in electronic structure calculations, while TTNS is more commonly employed in quantum dynamics simulations. These findings are consistent with previous reports~\cite{larsson2024tensor,gunst2018t3ns, li2021expressibility,larsson2019computing}.
}

\subsection{Spin Relaxation Dynamics of the Spin-Boson Model}
\label{sec:sbm}
We next apply our algorithm to the spin relaxation dynamics of the spin-boson model using the Cole-Davidson spectral density~\cite{zhou2012dynamics}
\begin{equation}
\label{eq:sdf-cd}
\begin{aligned}
    J_{\textrm{CD}}(\omega) & = \eta \frac{\sin{\betacd \thetacd}}{\left [ 1+(\omega^2/\omega_c^2) \right]^{\betacd/2}} \\ 
    \thetacd & = \arctan{\frac{\omega}{\omega_c}} \ ,
\end{aligned}
\end{equation}
where $\betacd < 1$ is the fractional stretching exponent.
The Cole-Davidson spectral density function exhibits distinct behavior at low frequency and high frequency limits.
When $\omega \ll \omega_c$, $J_{\textrm{CD}} \approx \eta \betacd \frac{\omega}{\omega_c}$.
When $\omega \gg \omega_c$, $J_{\textrm{CD}} \approx \eta \sin{\left(\frac{\pi \betacd}{2}\right)} \left (\frac{\omega_c}{\omega}\right)^{\betacd}$ which shows slow power law decay.
Despite the complicated form of $J_{\textrm{CD}}$, the reorganization energy $\lambda$ is simply related to $\eta$ by $\lambda=2\eta$.
The key difference between $J_{\textrm{CD}}$ and $J_{\textrm{Ohmic}}$ is that $J_{\textrm{CD}}$ has a long tail at the high-frequency region, particularly when $\betacd$ is small.
The contribution from the high frequency part can be taken into account through a Born-Oppenheimer type approximation, resulting in a modified $\Delta$
\begin{equation}
    \Delta^{\textrm{eff}} = \Delta \exp{-\frac{2}{\pi} \int_{\omega_q}^\infty \frac{J_{\textrm{CD}}(\omega)}{\omega^2} d\omega} \ ,
\end{equation}
where $\omega_q$ is the cutoff frequency during mode discretization.
The density of states for mode discretization is chosen as
\begin{equation}
    \rho(\omega) \propto \frac{J_{\textrm{CD}}(\omega)}{\omega} \ .
\end{equation}
The other setups, unless otherwise specified, are the same as those in Sec.~\ref{sec:scaling}.
A binary tree as shown in Fig.~\ref{fig:tree_structure}(b) is used for TTNS topology.
We set $\Ms=20$ and $N_b=1000$.
For low-frequency modes, $d$ ranges from dozens to hundreds, and for high-frequency modes $d$ is chosen as 4.
We first validate our algorithm by reproducing previous ML-MCTDH results,
shown in Fig.~\ref{fig:cole_davidson} in the Appendix.

The finite temperature dynamics of the spin relaxation is obtained through thermo field dynamics~\cite{takahashi1996thermo, borrelli2016quantum, borrelli2021finite, Peter21}.
In this method, the finite temperature density matrix of the bath environment is transformed into a pure state by introducing an auxiliary space $Q$, analogous  to the physical space $P$.
The finite temperature dynamics is then reduced to the zero temperature dynamics with a transformed Hamiltonian
\begin{equation}
\label{eq:h-tfd-phonon}
    \hat {\bar H} =   \Delta \hat \sigma^x  + \frac{1}{2}\sum_{i} (\hat p^2_i + \omega_i^2\hat q_i^2) - 
    \frac{1}{2}\sum_{i} (\hat{\tilde{p}}^2_i + \omega_i^2\hat{\tilde{q}}_i^2) 
    + \hat \sigma^z \sum_i  c_i \cosh{\theta_i}  \hat q_i +  \hat \sigma^z\sum_i   c_i \sinh{\theta_i} \hat{\tilde{q}}_i \ ,
\end{equation}
where operators with a tilde ``$\tilde{ \ }$'' are operators in the $Q$ space and $\theta_i = \arctanh{\exp{-\frac{\omega_i}{2 k_B T}}}$.
The initial thermal state of the bath is represented by $\ket{0}_P \ket{0}_Q$ in the harmonic oscillator eigenbasis.

\begin{figure}[H]
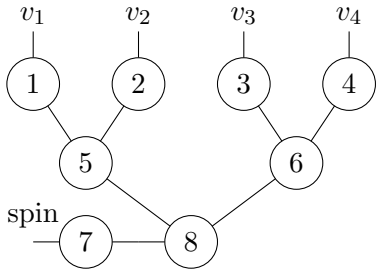

    \centering

    \begin{diagram}
    
    \draw (4,-1.5) -- (2,0);
    \draw (4,-1.5) -- (6,0);

    \MpsCircle{2}{-1.5}{7}{lr};
    \MpsCircle{4}{-1.5}{8}{l};
    \draw (1, -1) node {spin};
    
    \draw (2,0) -- (1,1.5);
    \draw (2,0) -- (3,1.5);
    \draw (6,0) -- (5,1.5);
    \draw (6,0) -- (7,1.5);

    \MpsCircle{2}{0}{5}{};
    \MpsCircle{6}{0}{6}{};

    \MpsCircle{1}{1.5}{1}{u};
    \MpsCircle{3}{1.5}{2}{u};    
    \MpsCircle{5}{1.5}{3}{u};
    \MpsCircle{7}{1.5}{4}{u};  
    
    \draw (1, 2.8) node {$v_1$};
    \draw (3, 2.8) node {$v_2$};
    \draw (5, 2.8) node {$v_3$};
    \draw (7, 2.8) node {$v_4$};
    
    \draw (4,0) node (X){};
    
    \end{diagram}  

    \caption{\rev{Schematic diagram for the tree nodes in the spin-boson model. Each node is labeled with an index.}}
    \label{fig:tree_sbm}
\end{figure}

\rev{
We next provide more details on the automatically constructed TTNO for the spin-boson model. The tree tensor network follows a binary tree topology, as depicted in Fig 4(b) and reproduced in Fig.~\ref{fig:tree_sbm}. The tree only shows 4 vibration modes for simplicity, but our actual simulation uses 1000 modes. For reference, each node is labeled with an index.
}

\rev{
We then show the symbolic tensors $\hat W[i]$ for the corresponding TTNO.
For the leaf nodes ($i =1, 2, 3, 4$), $\hat W[i]$ has only one index
\begin{equation}
    \hat W[i] = \mqty[\hat I_i \\ 
    \frac{1}{2}\hat p^2_i + \frac{1}{2}\omega_i^2 \hat q^2_i \\ 
    c_i \hat q_i], \ i =1,2,3,4 \ .
\end{equation}

For the body nodes ($i =5, 6$), $\hat W[i]$ has three indices and the shape is $(3, 3, 3)$. The first two indices connect to the children and the last index connects to the parent. Since the body nodes are not associated with any physical degree of freedom, the possible matrix elements for $\hat W[i]$ are 0 and 1 and $\hat W[i]$ is constructed as
\begin{equation}
    \hat W[i]_{jk} = \mqty[\delta_{j1}\delta_{k1} \\
    \delta_{j1}\delta_{k2} + \delta_{j2}\delta_{k1} \\
    \delta_{j1}\delta_{k3} + \delta_{j3}\delta_{k1}], \ i =5, 6 \ .
\end{equation}
Here $j$ and $k$ are indices to the children and the index to the parent is shown as the vector.
The vector elements correspond to the identity operator, the vibration energy and the vibration coordinate for the vibrations respectively.

The local TTNO tensor $\hat W$ for the spin node is
\begin{equation}
    \hat W[7] = \mqty[\hat \sigma^z \\
    \hat \sigma^x \\
    \hat I] \ .
\end{equation}

The final virtual node with $i=8$ has three indices $jkl$, connecting to nodes 5, 6, 7, respectively. $\hat W[8]$ is then
\begin{equation}
    \hat W[8]_{jkl} = 
      \delta_{j1} \delta_{k1}\delta_{l2}
    + \delta_{j1} \delta_{k3}\delta_{l1}
    + \delta_{j3} \delta_{k1}\delta_{l1}
    + \delta_{j1} \delta_{k2}\delta_{l3}
    + \delta_{j2} \delta_{k1}\delta_{l3}
\end{equation}

Regarding the treatment of the auxiliary space for finite temperature dynamics simulation, 
the transformed Hamiltonian Eq. 21 has the same form as the spin-boson Hamiltonian Eq. 13, despite that the number of vibration modes is doubled.
Consequently, the constructed TTNO at finite temperature is identical to the TTNO at zero temperature.
Because the computational cost scales linearly with the number of modes in the spin-boson model, we conclude that the finite temperature computation is twice as costly as zero temperature computation. 
}

\begin{figure}[htpb]
    \centering    \includegraphics[width=\textwidth]{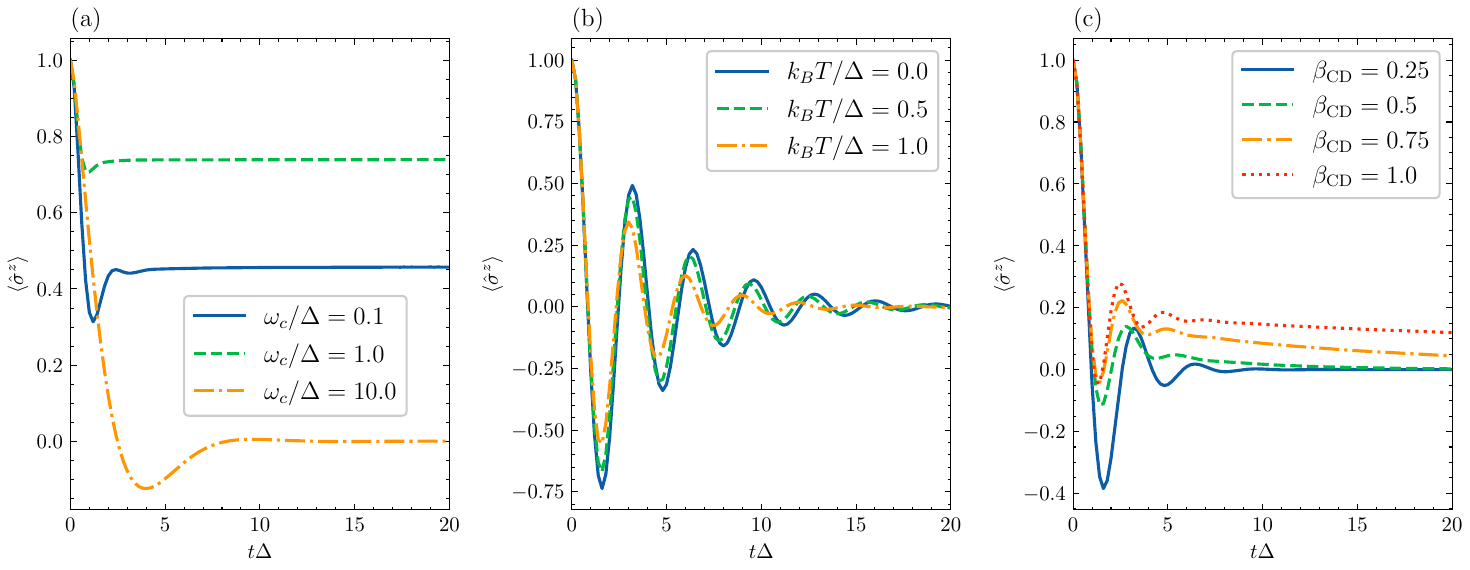}
    \caption{Spin relaxation dynamics based on the Cole-Davidson spectral density. (a)$\eta/\Delta=10$, $\betacd=0.5$, and $k_BT/\Delta =0$. (b) $\eta/\Delta=1$, $\betacd=0.25$ and $\omega_c/\Delta=1$. (c) $\eta/\Delta=2.5$, $\omega_c/\Delta=1$ and $k_BT/\Delta =0$.}
    \label{fig:cd_production}
\end{figure}

Fig.~\ref{fig:cd_production} shows the simulated spin relaxation dynamics using the Cole-Davidson spectral density.
In Fig.~\ref{fig:cd_production}(a), we investigate the influence of the characteristic frequency $\omega_c$ 
on the spin relaxation dynamics, with $\eta/\Delta=10$  and $\betacd=0.5$.
Notably, when the vibration frequency and tunneling constant are comparable (i.e., $\omega_c/\Delta=1$), the spin relaxation exhibits the most incoherent behavior compared to other cases. 
Additionally, due to the strong system-bath coupling,  $\braket{\hat \sigma^z}$ demonstrates localization in both the adiabatic and the intermediate regimes.
We then explore the impact of temperature on the dynamics in Fig.~\ref{fig:cd_production}(b),
with $\eta/\Delta=1$, $\betacd=0.25$ and $\omega_c/\Delta=1$.
As the temperature increases, the spin dynamics become increasingly incoherent.
For this figure, we employ $N_b=500$ including both $P$ and $Q$ space and $\Ms=96$ due to the strong entanglement at finite temperature.
Lastly, we study the effect of $\betacd$ on the spin relaxation dynamics.
As shown in Fig.~\ref{fig:cd_production}(c), the dynamics of the spin becomes more incoherent with higher values of $\betacd$.
The other parameters are $\eta/\Delta=2.5$ and $\omega_c/\Delta=1$.
We note that a lower value of $\betacd$ enhances the contribution of high-frequency vibration modes.
Up to $t\Delta=20$, localization is not observed for the values of $\betacd$ considered.

\subsection{Charge Transport in a Molecular Junction}
\label{sec:transport}
We then move on to study quantum charge transport in a molecule junction with vibrational coupling.
Single molecular junction has been
widely used to investigate nonequilibrium many-body quantum phenomena at the nanoscale~\cite{reed1997conductance, chen2007measurement, galperin2008nuclear, xiang2016molecular}.
The model for vibrationally coupled electron transport through a
single-molecule junction can usually be divided into three parts~\cite{wang2013multilayer,wang2013numerically, wang2016employing, yang2023time}.
The first part is 
the electronic part, which includes one discrete electronic
state at the molecular bridge and two identical electronic continua
describing the left and the right metal leads
\begin{equation}
    \hat H_{\textrm{el}} = E_d \hat d^\dagger \hat d + \sum_{\zeta =L, R}\sum_{k_\zeta} E_{k_\zeta} \hat c^\dagger_{k_\zeta} \hat c_{k_\zeta} + \sum_{\zeta=L, R}\sum_{k_\zeta} V_{dk_\zeta} \left (\hat d^\dagger \hat c_{k_\zeta} + \hat c^\dagger_{k_\zeta} \hat d \right) \ .
\end{equation}
Here, $\hat d^\dagger$($\hat d$) and $\hat c^\dagger$($\hat c$) are fermionic creation (annihilation) operators for the electronic state on the molecular bridge and the metal leads respectively.
The electronic states in the left (right) lead state are indexed with $k_L$ and $k_R$ respectively.
Note that the partition between left and right leads should not be confused with the partition in the bipartite graph in Sec.~\ref{sec:algo-ttno}.
$E_d$ is the site energy of the molecular bridge and is set to 0 throughout this section.
$E_{k_\zeta}$ is the energy of the lead states and $V_{dk_\zeta}$ is the molecule-lead coupling strength, which will be provided later.
Since $\hat d^\dagger$($\hat d$) and $\hat c^\dagger$($\hat c$) have to follow the anti-commutation  property of fermionic operators, they are transformed to spin operators through the Jordan-Wigner transformation~\cite{JW28}
\begin{equation}
\begin{aligned}
    \hat a^\dagger_k = \prod_{j=1}^{k-1} \hat \sigma^z_j \hat \sigma^+_k \ , \\
    \hat a_k = \prod_{j=1}^{k-1} \hat \sigma^z_j \hat \sigma^-_k \ , \\    
\end{aligned}
\end{equation}
where $\hat \sigma^+=\frac{1}{2}(\hat \sigma^x-i\hat\sigma^y)$ and $\hat \sigma^-=\frac{1}{2}(\hat \sigma^x+i\hat\sigma^y)$ are Pauli ladder operators.

The molecular bridge is then coupled to a phonon bath
\begin{equation}
\begin{aligned}
 \hat H_{\textrm{el-nuc}} & = \hat d^\dagger \hat d \sum_j 2 c_j \hat q_j \ , \\
 \hat H_{\textrm{nuc}} & =  \frac{1}{2}\sum_j (\hat p^2_j + \omega_j^2\hat q_j^2) \ .
\end{aligned}
\end{equation}
The total Hamiltonian is then written as
\begin{equation}
    \hat H = \hat H_{\textrm{el}} + \hat H_{\textrm{el-nuc}} + \hat H_{\textrm{nuc}} \ .
\end{equation}

The primary physical observable for quantum transport through a molecular junction is the electronic current for a given source-drain bias voltage.
The electronic current operator for each lead reads
\begin{equation}
    \hat I_\zeta = i[\hat H, \hat N_\zeta] = i \sum_{k_\zeta} V_{dk_\zeta} \left (\hat d^\dagger \hat c_{k_\zeta} - \hat c^\dagger_{k_\zeta} \hat d \right ) \ , \quad \zeta = L, R \ ,
\end{equation}
where $\hat N_\zeta=\sum_{k_\zeta} \hat c^\dagger_{k_\zeta} \hat c_{k_\zeta}$ is the occupation number operator for each lead.
The overall time-dependent current $I(t)$ is calculated with
\begin{equation}
    I(t) = \frac{1}{2} \left [\braket{\hat I_R(t)} - \braket{\hat I_L(t)} \right] \ .
\end{equation}

The source-drain bias voltage is considered through different initial lead states based on the grand-canonical ensemble.
More specifically, the initial density matrix for lead $\zeta$ is
\begin{equation}
    \rho_\zeta = \exp{-\sum_{k_\zeta} (E_{k_\zeta} -\mu_\zeta)\hat c^\dagger_{k_\zeta} \hat c_{k_\zeta} /k_B T} \ ,
\end{equation}
where $\mu_\zeta$ is the chemical potential for lead $\zeta$, given by
\begin{equation}
    \mu_{L/R}=\pm V/2 \ ,
\end{equation}
where $V$ is the source-drain bias voltage.
Furthermore, we consider two different initial states of the molecular bridge: occupied and unoccupied.
In both cases, the oscillator bath is in equilibrium with the state of the molecule bridge.
When the molecular bridge is occupied, the coordinate operator $\hat q$ of the phonon modes
is replaced with another displaced coordinate operator $\hat Q$ according to $\hat q = \hat Q - 2c_j/\omega_j^2$.

Next, we turn to the parameters in the charge transport Hamiltonian.
The electronic energies $E_{k_\zeta}$ and molecule-lead coupling strengths $V_{dk_\zeta}$ are defined through energy-dependent level width functions
\begin{equation}
    \Gamma_\zeta(E)=2\pi \sum_{k_\zeta} |V_{dk_\zeta}|^2 \delta(E-E_{k_\zeta}) \ ,
\end{equation}
which is analogous  to the spectral density function for the phonon bath defined in Eq.~\ref{eq:sdf}.
In this work we choose a tight-binding model for $\Gamma_\zeta$
\begin{equation}
    \Gamma(E) = 
    \begin{cases}
        \frac{\alpha_e^2}{\beta_e^2}\sqrt{4\beta_e^2-E^2} \ , \quad |E| < 2|\beta_e|\\
        0\ , \quad |E| > 2|\beta_e|
    \end{cases} \ ,
\end{equation}
where $\beta_e$ and $\alpha_e$ are nearest-neighbour coupling between two lead sites and between the lead site and the bridge state, respectively.
$\Gamma$ is then discretized according to Eq.~\ref{eq:discre} to produce the lead site energy $E_{k_\zeta}$ and the coupling constant $V_{dk_{\zeta}}$.
In this work, we use a constant density of state, leading to an equidistant discretization of the interval $[-2\beta_e, 2\beta_e]$.
We set $\alpha_e=0.2$ eV, $\beta_e=1$ eV, and a bias voltage of $0.1$ V.
For each lead, 160 electronic states are discretized.
The phonon bath is modeled using the Cole-Davidson spectral density and the prescription to obtain the discrete modes is the same as that described in Sec.~\ref{sec:sbm}.
We set $\eta=1000 \  \cminv{}$, $\omega_c=500 \  \cminv{}$ and $\betacd=0.5$.
The number of bath modes $N_b$ is 1000.
The overall tree structure for the model is based on the binary tree in Fig.~\ref{fig:tree_structure}(b).
Two binary trees are constructed for lead states with $E_{k_{\zeta}} - \mu_\zeta < E_d$ 
and $E_{k_{\zeta}} - \mu_\zeta > E_d$ respectively~\cite{rams2020breaking, yang2023time}.
The two trees are then grouped together for an overall binary tree.
Another binary tree is constructed for the phonon bath.
The two binary trees are then attached to the root, which represents the molecular bridge.

Similar to the study of the spin-boson model, the finite temperature time-dependent current is calculated through thermo field dynamics~\cite{takahashi1996thermo, yang2023time}.
The transformed Hamiltonian for the phonon bath and the electron-phonon coupling is the same as Eq.~\ref{eq:h-tfd-phonon}.
Meanwhile, the electronic part is transformed as follows
\begin{equation}
\begin{aligned}
    \hat{\bar{H}}_{\textrm{el}} & = E_d \hat d^\dagger \hat d + \sum_{\zeta =L, R}\sum_{k_\zeta} E_{k_\zeta} \left ( \hat c^\dagger_{k_\zeta} \hat c_{k_\zeta} - \hat{\tilde{c}}^\dagger_{k_\zeta} \hat{\tilde{c}}_{k_\zeta} \right ) \\
    & \quad + \sum_{\zeta=L, R}\sum_{k_\zeta} V_{dk_\zeta} \left (\cos{\theta_{k_\zeta}}\hat d^\dagger \hat c_{k_\zeta} + \cos{\theta_{k_\zeta}} \hat c^\dagger_{k_\zeta} \hat d + \sin{\theta_{k_\zeta}}\hat d^\dagger \hat{\tilde{c}}^\dagger_{k_\zeta}
    + \sin{\theta_{k_\zeta}} \hat{\tilde{c}}_{k_\zeta} \hat d \right) \ ,
\end{aligned}
\end{equation}
where $\theta_{k_\zeta} = \arctan{\exp{-\frac{E_{k_\zeta}-\mu_{\zeta}}{2 k_B T}}}$.
The transformation over the current operator follows similarly
\begin{equation}
     \hat I_\zeta = i \sum_{k_\zeta} V_{dk_\zeta} \left (\cos{\theta_{k_\zeta}} \hat d^\dagger \hat c_{k_\zeta} - \cos{\theta_{k_\zeta}} \hat c^\dagger_{k_\zeta} \hat d 
     + \sin{\theta_{k_\zeta}} \hat{\tilde{c}}_{k_\zeta} \hat d 
     - \sin{\theta_{k_\zeta}} \hat d^\dagger \hat{\tilde{c}}^\dagger_{k_\zeta}
     \right) \ , \quad \zeta = L, R \ .
\end{equation}
In this section, the time-dependent currents are calculated with 300 K unless otherwise specified.

The overall formalism of the quantum transport problem appears to be much more complicated than the spin-boson model.
However, thanks to the automatic construction of TTNO, both the programming effort and the computational cost for the simulation are of the same order as the simulation of spin-boson model.
For example, the overall Python script for the simulation of the quantum transport problem contains approximately 300 lines of code, 
whereas the script for the simulation of the spin-boson model at zero temperature has around 200 lines.
These scripts handle tasks such as the determination of model parameters through the discretization  of the spectral density function, construction of the SOP Hamiltonian and current operators, specification of the TTN tree structure, construction of the TTNOs, time evolution, the calculation of physical observables, and various logging outputs.
In these scripts we only rely on library functions and classes 
that are general and applicable to any other physical model. 
Additionally, the maximum $\Mo$ across all edges for the TTNO of the transformed Hamiltonian is 5.
This ensures linear scaling with respect to both the number of electron modes and the number of phonon modes in the model.

\rev{
We next provide more details on the automatically constructed TTNO for the molecular junction model. 
First of all, the overall tree structure is depicted in Fig.~\ref{fig:tree_junction_overall}.
Two binary trees are constructed for lead states with $E_{k_{\zeta}} - \mu_\zeta < E_d$ 
and $E_{k_{\zeta}} - \mu_\zeta > E_d$ respectively.
The two trees are then grouped together for an overall binary tree.
Another binary tree is constructed for the phonon bath.
The two binary trees are then attached to the root, which represents the molecular bridge.
\begin{figure}[H]
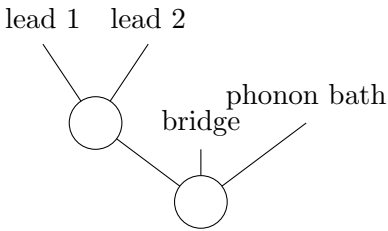

    \centering  
    \begin{diagram}
    
    \draw (4,-1.5) -- (2,0);
    \draw (4,-1.5) -- (6,0);

    \MpsCircle{4}{-1.5}{}{u};
    \draw (4, 0) node {bridge};
    
    \draw (2,0) -- (1,1.5);
    \draw (2,0) -- (3,1.5);
    \draw (6, 0.5) node {phonon bath};

    \MpsCircle{2}{0}{}{}; 
    
    \draw (1, 2) node {lead 1};
    \draw (3, 2) node {lead 2};
    
    \draw (4,0) node (X){};
    
    \end{diagram}  
    \caption{\rev{Schematic diagram for the tree structure of the molecular junction model.}}
    \label{fig:tree_junction_overall}
\end{figure}

The combined system of the bridge and the phonon bath is similar to a spin-boson model.
As a result, the TTNO structure of the phonon bath tree is the same for both the molecular junction model and the spin-boson model.
The maximum bond dimension $\Mo$ in this part of the tree is then 3.

For the electronic lead, the maximum bond dimension $\Mo$ is 5.
Since each electronic mode has a physical bond dimension of only 2, two modes are attached directly to one node without primitive basis contraction.
In Fig.~\ref{fig:tree_junction_lead}, we depict a schematic diagram of the tree tensor network corresponding to the electronic leads.
Here for ease of demonstration, only 8 lead electronic modes are shown.

\begin{figure}[H]
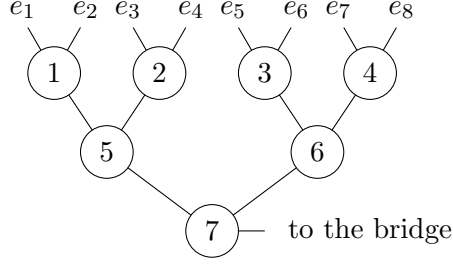

    \centering   
    \begin{diagram}
    
    \draw (4,-1.5) -- (2,0);
    \draw (4,-1.5) -- (6,0);

    \MpsCircle{4}{-1.5}{7}{r};
    \draw (7, -1.5) node {to the bridge};
    
    \draw (2,0) -- (1,1.5);
    \draw (2,0) -- (3,1.5);
    \draw (6,0) -- (5,1.5);
    \draw (6,0) -- (7,1.5);

    \MpsCircle{2}{0}{5}{};
    \MpsCircle{6}{0}{6}{};

    \MpsCircle{1}{1.5}{1}{b};
    \MpsCircle{3}{1.5}{2}{b};    
    \MpsCircle{5}{1.5}{3}{b};
    \MpsCircle{7}{1.5}{4}{b};  
    
    \draw (0.4, 2.7) node {$e_1$};
    \draw (1.6, 2.7) node {$e_2$};
    \draw (2.4, 2.7) node {$e_3$};
    \draw (3.6, 2.7) node {$e_4$};
    \draw (4.4, 2.7) node {$e_5$};
    \draw (5.6, 2.7) node {$e_6$};
    \draw (6.4, 2.7) node {$e_7$};
    \draw (7.6, 2.7) node {$e_8$};
    
    \draw (4,0) node (X){};
    
    \end{diagram}  

    \caption{\rev{Schematic diagram for the tree nodes of the lead part in the molecular junction model. $e_i$ represents the $i$th degree of freedom of the two leads. Each node is labeled with an index. The molecular bridge and the phonon bath are omitted.}}
    \label{fig:tree_junction_lead}
\end{figure}

The TTNO for the molecular junction is more complex than the TTNO for the spin-boson model.
For nodes 1 and 4, the symbolic TTNO tensor has one index and the dimension is 4
\begin{equation}
\label{eq:ttno_junction1}
    \hat W[i] = \mqty[\hat I_{2i-1} \hat I_{2i} \\
    E_{2i-1} \hat \sigma^+_{2i-1} \hat \sigma^-_{2i-1} + E_{2i} \hat \sigma^+_{2i} \hat \sigma^-_{2i}  \\
    V_{2i-1} \hat \sigma^+_{2i-1} \hat \sigma^z_{2i} + V_{2i} \hat I_{2i-1} \hat \sigma^+_{2i}  \\
    V_{2i-1} \hat \sigma^-_{2i-1} \hat \sigma^z_{2i} + V_{2i} \hat I_{2i-1} \hat \sigma^-_{2i} ] , \ i=1, 4 \ .
\end{equation}
Note that here spin operators are employed due to Jordan-Wigner transformation.

For nodes 2 and 3, the symbolic TTNO tensor has one index and the dimension is 5, which is the maximum bond dimension in the whole TTNO tree
\begin{equation}
\label{eq:ttno_junction2}
    \hat W[i] = \mqty[\hat I_{2i-1} \hat I_{2i} \\
    \hat \sigma^z_{2i-1} \hat \sigma^z_{2i} \\
    E_{2i-1} \hat \sigma^+_{2i-1} \hat \sigma^-_{2i-1} + E_{2i} \hat \sigma^+_{2i} \hat \sigma^-_{2i}  \\
    V_{2i-1} \hat \sigma^+_{2i-1} \hat \sigma^z_{2i} + V_{2i} \hat I_{2i-1} \hat \sigma^+_{2i}  \\
    V_{2i-1} \hat \sigma^-_{2i-1} \hat \sigma^z_{2i} + V_{2i} \hat I_{2i-1} \hat \sigma^-_{2i} ] , \ i=2, 3 \ .
\end{equation}

The additional element in Eq.~\ref{eq:ttno_junction2} compared with Eq.~\ref{eq:ttno_junction1} is $\hat \sigma^z_{2i-1} \hat \sigma^z_{2i}$, which is the result of Jordan-Wigner transformation. 
The nodes at the ``boundary'' of the tree, such as nodes 1 and 4, do not have this term. 

The nodes 5 and 6 are three-indexed tensors and the shapes are (4, 5, 4) and (5, 4, 4) respectively. Taking the 5th node as an example, the local TTNO tensor is
\begin{equation}
\label{eq:ttno-nature}
    \hat W[5]_{jk} = \mqty[
    \delta_{j1}\delta_{k1} \\
    \delta_{j1}\delta_{k3} +\delta_{j2}\delta_{k1} \\
    \delta_{j1}\delta_{k4} + \delta_{j3}\delta_{k2} \\
    \delta_{j1}\delta_{k5} + \delta_{j4}\delta_{k2}
    ]
\end{equation}
The elements represent the identity operator, lead site energy, creation operator from the lead, and annihilation operator from the lead, respectively.

The 7th node, or the root node of the lead part, has 3 indices and the shape is (4, 4, 4).
The local TTNO tensor is
\begin{equation}
    \hat W[7]_{jk} = \mqty[
    \delta_{j1}\delta_{k1} \\
    \delta_{j1}\delta_{k2} +  \delta_{j2}\delta_{k1}\\
    \delta_{j1}\delta_{k3} +  \delta_{j3}\delta_{k1}\\
    \delta_{j1}\delta_{k4} +  \delta_{j4}\delta_{k1}\\
    ]
\end{equation}
The nature of the elements is the same as Eq.~\ref{eq:ttno-nature} and the role of the 7th node is to merge the operators from two sub-trees.

Lastly, the symbolic TTNO tensor for the bridge node is
\begin{equation}
    \hat W[\textrm{bridge}] = \mqty[
    E_d \hat \sigma^+_d\sigma^-_d & \hat I_d & 2\hat \sigma^+_d\sigma^-_d \\
    \hat I_d & 0 & 0  \\
    \hat \sigma^-_d & 0 & 0  \\
    \hat \sigma^+_d & 0 & 0 
    ]
\end{equation}

As in the spin-boson model, the finite temperature effect in the molecular junction model is taken into account via thermo field dynamicss~\cite{takahashi1996thermo}. The transformed Hamiltonian is structurally similar to the original Hamiltonian.
Thus, the structure of the constructed TTNO is the same.
}

\begin{figure}[htpb]
    \centering    \includegraphics[width=\textwidth]{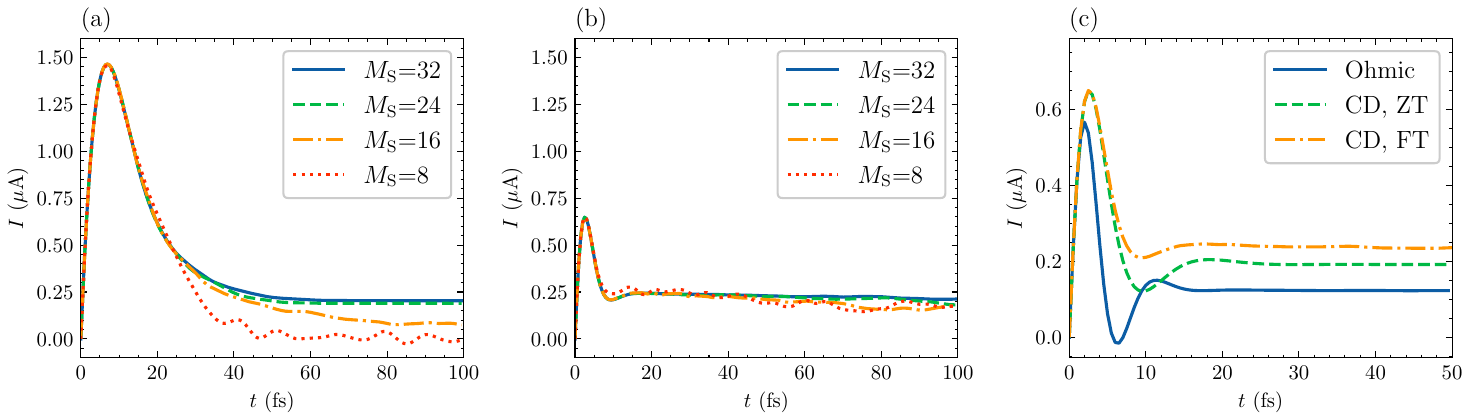}
    \caption{Time-dependent current for the charge transport in a molecular junction. The Cole-Davidson spectral density function is employed and the temperature is set to 300 K. In (a) and (b) the initial state of the molecular bridge is unoccupied and occupied respectively. In (c) the finite temperature (FT) result is compared with zero temperature (ZT) result with both Cole-Davidson (CD) spectral density and Ohmic spectral density.}
    \label{fig:transport}
\end{figure}

The simulated time-dependent current for the charge transport model is illustrated in Fig.~\ref{fig:transport}.
Fig.~\ref{fig:transport}(a) shows the convergence with respect to $\Ms$ when the molecular bridge is initially unoccupied.
The regularization technique from the references~\cite{wang2013multilayer, wang2013numerically} is used when $\Ms=24$ and $\Ms=32$ to achieve a converged current.
Fig.~\ref{fig:transport}(b) demonstrates the convergence with respect to $\Ms$ when the molecular bridge is initially occupied.
In both cases, $\Ms=32$ is sufficient to produce a converged outcome.
Interestingly, an initially unoccupied bridge leads to a larger transient current, which is consistent with previous studies.
Despite the different initial states, the steady current in Fig.~\ref{fig:transport}(a) and Fig.~\ref{fig:transport}(b) appears to be the same.
We further compare the finite-temperature current with zero-temperature current and zero-temperature current with Ohmic spectral density in Fig.~\ref{fig:transport}(c).
For the Ohmic spectral density, we use the same characteristic frequency $\omega_c$ and reorganization energy $\lambda$ as the Cole-Davidson spectral density.
From Fig.~\ref{fig:transport}(c) it appears that using Cole-Davidson spectral density leads to a higher steady current. Furthermore, increasing temperature also results in a higher steady current.

\section{Conclusion and Outlook}

In this work, we present an efficient implementation of the TTNS algorithm,
which is included as a module of the open-source Python package \textsc{Renormalizer}.
Our implementation features the automatic construction of the optimal TTNO based on bipartite graph theory.
For the models studied in this work, including the spin-boson model and the charge transport model, the TTNO of the Hamiltonian has constant $\Mo$.
Consequently, our algorithm scales linearly with the number of modes in the model, given fixed $\Ms$.
It is important to note that our algorithm for the construction of TTNO is completely general. 
It can be applied to any SOP operator and any tensor network with a tree topology.
Therefore, the application of our program is not limited to the simulation of open quantum systems.
It can also be applied to the simulation of other physical models, the propagation of stochastic Schrodinger equation~\cite{gao2022non}, and the simulation of quantum circuits~\cite{seitz2023simulating}.
While our program has broad applications, there is still plenty of room for further software improvement.
For example, the efficiency of our program can be improved through massive parallelism on both CPU and GPU~\cite{secular2020parallel} or mixed-precision computation.
In our previous work, we have shown that using a GPU can accelerate the time evolution of TD-DMRG by dozens of times~\cite{li2020numerical}.
Furthermore, just-in-time compilation techniques, which have been widely used to accelerate neural network training and the simulation of quantum circuits, could potentially be used to speed up the computational bottleneck of TTNS algorithms, such as the contraction of  $\hat H^{\textrm{eff}}_i A[i]$.

\begin{acknowledgements}
Weitang Li is supported by the Young Elite Scientists Sponsorship Program by CAST, 2023QNRC001.
Zhigang Shuai is supported by the National Natural Science Foundation of China (Grant No. T2350009) and the Guangdong Provincial Natural Science Foundation (Grant No. 2024A1515011185), as well as the Shenzhen city ``Pengcheng Peacock'' Talent Program. Jiajun Ren is supported by the National Natural
Science Foundation of China (Grant No. 22273005).
\end{acknowledgements}

\section*{Conflict of interest}
The authors have no conflicts to disclose.

\rev{
\section*{Data Availability}
The data that support the findings of this study are available from \url{https://github.com/liwt31/li2024optimal}.

\section*{Code Availability}
The code for this study is available from 
\url{https://github.com/liwt31/li2024optimal} and 
\url{https://github.com/shuaigroup/Renormalizer}.
}

\bibliography{refs}

\appendix
\section{Benchmark with ML-MCTDH}
\begin{figure}[htpb]
    \centering    \includegraphics[width=0.7\textwidth]{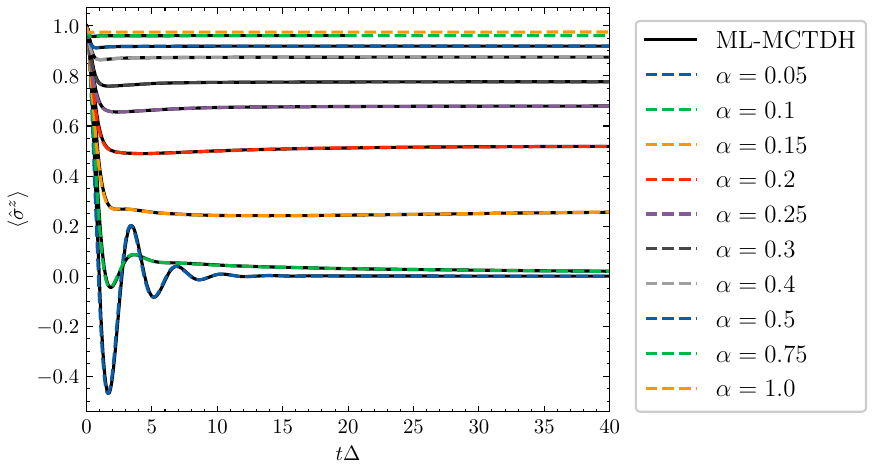}
    \caption{Spin relaxation dynamics of the spin-boson model with sub-Ohmic spectral density from weak coupling $\alpha=0.05$ to strong coupling $\alpha=1.0$. The colored dashed lines are the results of our TTNS implementaion with $\Ms = 20, d = 10$. The black solid lines are the results of ML-MCTDH with $\alpha$ from 0.05 to 0.75.}
    \label{fig:subohmic}
\end{figure}

In this section, we benchmark our implementation by reproducing existing results from ML-MCTDH, demonstrating the accuracy of our implementation. 
Firstly,
we consider the spin-boson model with the sub-Ohmic spectral density function as specified by Eq.~\ref{eq:sdf-ohmic}.
$s = 0.5$ and $\omega_c =20\Delta$ are employed as the parameters for the spectral density.
We employ $N_b=1000$, $\Ms=20$, and $d=10$ as the parameters for the TTNS calculation.
The other setups are the same as those for Fig.~\ref{fig:wall_time}.
We vary the coupling strength $\alpha$ from the weak coupling regime ($\alpha=0.05$)
to the strong coupling regime ($\alpha=1.0$).
In Fig.~\ref{fig:subohmic} we compare the dynamics calculated using our program with the results by ML-MCTDH~\cite{wang2010coherent}.
The black solid lines are the results of ML-MCTDH, and the colored dashed lines are the results of our TTNS program.
Across the weak to strong coupling regimes, the results show excellent agreement.

\begin{figure}[htpb]
    \centering    \includegraphics[width=\textwidth]{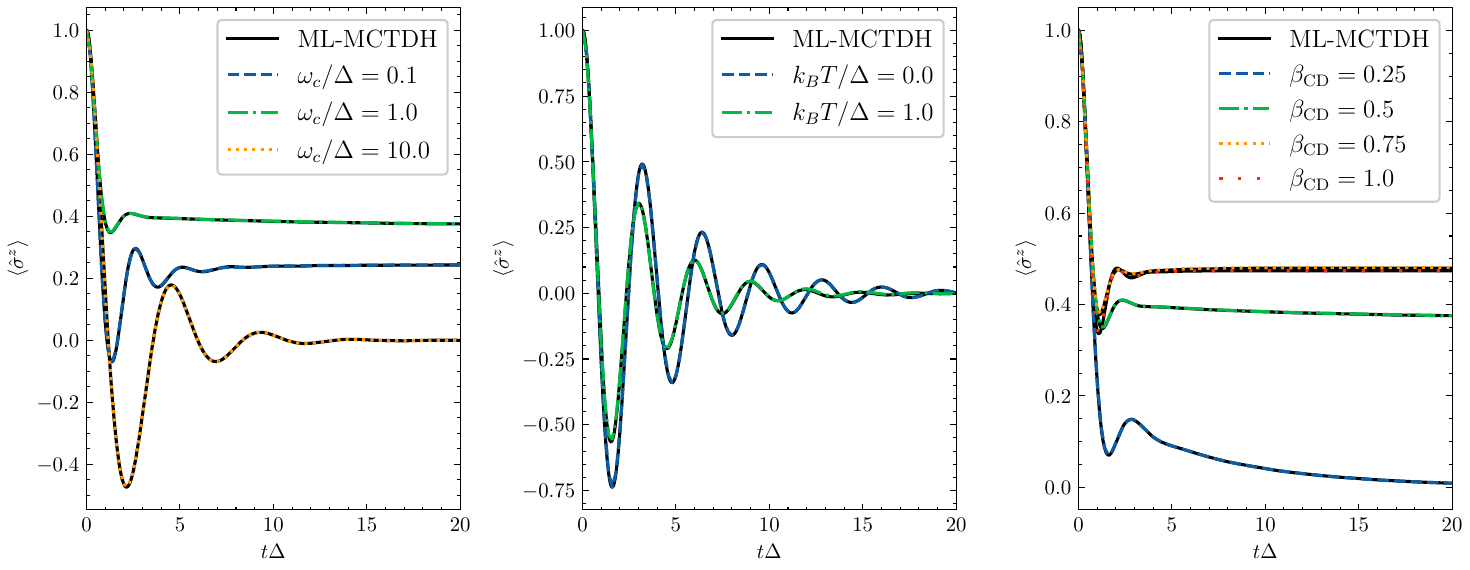}
    \caption{Spin relaxation dynamics based on the Cole-Davidson spectral density. (a)$\eta/\Delta=5$, $\betacd=0.5$, and $k_BT/\Delta =0$. (b) $\eta/\Delta=1$, $\betacd=0.25$ and $\omega_c/\Delta=1$. (c) $\eta/\Delta=5$, $\omega_c/\Delta=1$ and $k_BT/\Delta =0$.}
    \label{fig:cole_davidson}
\end{figure}

The second case involves the spin-boson model with the Cole-Davidson spectral density function as specified by Eq.~\ref{eq:sdf-cd}.
The computational setup is the same as those in Fig.~\ref{fig:cd_production}, but with different model parameters.
The details of the parameters are listed in the caption of Fig.~\ref{fig:cole_davidson}.
In Fig.~\ref{fig:cole_davidson}, the black solid lines are the results of ML-MCTDH, and the colored dashed lines are the results of our TTNS program.
Once again, the results are in excellent agreement, further validating the accuracy of our algorithm.

\end{document}